\documentclass[11pt]{article}
\usepackage{color,soul}
\usepackage{titlesec}
\usepackage[pagebackref]{hyperref}
\usepackage{dirtytalk}
\usepackage{lipsum}
\usepackage{ntheorem}
\usepackage{mathtools}
\usepackage{graphicx}
\usepackage{calc}
\usepackage{stmaryrd}
\usepackage{paralist}
\newlength{\depthofsumsign}
\usepackage{graphicx}
\usepackage{amsmath}
\usepackage{amssymb}
\usepackage{amsfonts}
\usepackage{array}
\usepackage{tikz}
\usepackage{hhline}
\usepackage{multirow}
\usepackage{subfig}
\usepackage{hyperref}
\usepackage{mathtools}
\usepackage{tikz}
\usepackage{makecell}
\usepackage{pgfplots}
\usepackage{afterpage}
\usepackage{fancyhdr}
\usepackage{setspace}
\usepackage{bm}
\usepackage[linesnumbered,ruled,vlined]{algorithm2e}
\SetKwInput{KwInput}{Input}
\SetKwInput{KwOutput}{Output}
\usepackage{authblk}

\DeclareMathOperator*{\argmax}{argmax}

\newtheorem{proposition} {Proposition}
\newtheorem*{proposition-non}{Proposition}

\newtheorem{remark}{Remark}
\newtheorem{example}{Example}

\allowdisplaybreaks

\newenvironment{proof}{\noindent{\bf Proof:}\indent}%
                      {\hfill $\Box$\par}

\newcommand{\sym}[1]{{\sf #1}}

\title{Use of Simple Arithmetic Operations to Construct Efficiently Implementable Boolean functions Possessing High Nonlinearity and Good Resistance to Algebraic Attacks}
\author[1,2]{Claude Carlet}
\author[3]{Palash Sarkar\thanks{Corresponding author.}}
\affil[1]{LAGA Laboratory, University of Paris 8, 93526 Saint-Denis, France}
\affil[2]{University of Bergen, Norway}
\affil[3]{Indian Statistical Institute, 203, B.T. Road, Kolkata, India 700108}
\affil[ ]{Emails: {\tt claude.carlet@gmail.com, palash@isical.ac.in}}

\date{\today}

\begin{document}

\maketitle              

\begin{abstract}
	We describe a new class of Boolean functions which provide the presently best known trade-off between low computational complexity, nonlinearity and
	(fast) algebraic immunity. In particular, for $n\leq 20$, we show that there are functions in the family achieving a combination of nonlinearity and 
	(fast) algebraic immunity which is superior to what is achieved by any other efficiently implementable function.
	The main novelty of our approach is to apply a judicious combination of simple integer and binary field arithmetic to Boolean function construction. 
\end{abstract}

\section{Introduction}
%Many cryptosystems, such as stream ciphers, use Boolean functions for providing what C. Shannon called confusion in~\cite{CC-Shannon49}. Concretely, confusion has been 
%specified in the nineties into a series of cryptographic criteria (see {\em e.g.}~\cite{BF-book}). 

The nonlinear filter model is a several decades old model for stream ciphers.
This model consists of two components, namely a linear feedback shift register (LFSR) and a Boolean function which is applied to a subset of the bits of the LFSR. 
The sequence of outputs of the Boolean function on the successive states of the LFSR constitutes the keystream produced by the 
stream cipher. 
%A basic requirement on the Boolean function is that it is balanced so that there is no statistical bias in the keystream produced by the stream cipher
%and so between the plain-text and the cipher-text. 

The mathematical challenge for a Boolean function to be used in the filter model of stream ciphers is the following. Construct a large family (if possible an infinite one)
of Boolean functions all of which are balanced and achieve a good combination of high nonlinearity and high algebraic resistance and further are efficient to implement. 
In~\cite{DBLP:journals/tit/Carlet22}, this design challenge was referred to as ``the big single-output Boolean problem'' 
(similar, in the domain of Boolean functions for stream ciphers, to the ``big APN problem'' in the domain of vectorial functions).
In more concrete terms, Section~3.1.5 of~\cite{BF-book} suggests that to resist algebraic attacks the number of variables should be at least 13, and further
goes on to recommend that in practice ``the number of variables will have to be near 20'' which then creates a ``problem of efficiency of the stream cipher.''

There are several known constructions of families of Boolean functions which achieve some, but not all of the above properties. We discuss these families in details
in Section~\ref{sec-prev}. For the present, we briefly mention some of these families. The Carlet-Feng (CF) functions~\cite{DBLP:conf/asiacrypt/CarletF08}
are balanced, achieve optimal algebraic immunity (and also almost optimal fast algebraic immunity) and high nonlinearity, but are not efficient to implement
(see~\cite{DBLP:journals/tit/Carlet11} for a gate count estimate of CF functions). 
The hidden weight bit (HWB) function~\cite{DBLP:journals/tc/Bryant91} is very efficient to implement and in~\cite{WCST} it was shown that the HWB function has good algebraic 
immunity, but the nonlinearity is too low. Subsequently, a sequence of works~\cite{DBLP:journals/amco/WangTS14,DBLP:journals/tit/Carlet22,MO24,cryptoeprint:2024/2022} 
have generalised the HWB function to improve the nonlinearity while retaining the properties of good algebraic immunity and being efficient to implement. The trade-offs achieved 
by these works are not completely satisfactory.

In this paper, we revisit the above mentioned mathematical challenge for Boolean functions. We describe a family of functions as a solution to the problem. 
The functions are based on the HWB function. To improve the nonlinearity, we introduce post-processing and pre-processing steps. For the post-processing step, we
first extend the HWB function to a vectorial function by extracting a few bits and then apply a highly nonlinear function to these bits. The number of extracted
bits is small (in fact, a constant) and so it is feasible to apply a highly nonlinear function to these bits without affecting the efficiency of implementation. 
For the pre-processing step, we design a novel bijection from $n$-bit strings to $n$-bit strings. The bijection is constructed by a judicious combination of 
simple integer and binary field arithmetic. 
To the best of our knowledge, no previous work reported construction of Boolean functions based on a combination of integer and binary field arithmetic. 
Since all operations that we use are simple and efficient, the overall construction is also quite efficient.

The net effect of applying both the pre and post processing steps is a significant improvement
of both nonlinearity and algebraic resistance over HWB without compromising on the issue of efficient implementation. Our experimental results show that for 
all $n\leq 20$, both nonlinearity and algebraic resistance of suitably chosen $n$-variable functions from the new family are substantially better than the corresponding values of 
$n$-variable functions from all previously known families~\cite{DBLP:journals/amco/WangTS14,DBLP:journals/tit/Carlet22,MO24,cryptoeprint:2024/2022} that are efficient to implement.
So our construction provides good solutions to the concrete problem highlighted in Section~3.1.5 of~\cite{BF-book}.

The paper is organised as follows. In Section~\ref{sec-prelim} we describe the preliminaries. 
The relevant previous constructions are discussed in Section~\ref{sec-prev}.
The family of functions is described in Section~\ref{sec-int-lambdaHWB}. Section~\ref{sec-conclu} concludes the paper. 
%\section{Preliminaries \label{sec-prelim}}
%\section{Relevant Previous Constructions \label{sec-prev}}
%\section{Construction of Interval $\lambda$-HWB Functions \label{sec-int-lambdaHWB}}
%\section{Conclusion \label{sec-conclu} }

%\begin{remark}\label{rem-code}
%	We report a number of experimental results. We used simple (and non-optimised) C code to construct the functions and compute their nonlinearities and
%	algebraic degrees. For computing algebraic immunity we used the Boolean function 
%	library\footnote{\url{https://doc.sagemath.org/html/en/reference/cryptography/sage/crypto/boolean_function.html\#sage.crypto.boolean_function.BooleanFunction.annihilator}} 
%	of the SageMath software. For computing fast algebraic immunity, we used a program written by Simon Fischer which was kindly provided to us by Deng Tang.
%\end{remark}

%}

\section{Preliminaries \label{sec-prelim}}
In this section, we introduce the notation and provide the definitions of the properties of Boolean functions that we consider in this work. For further
details and more elaborate discussion on these issues we refer to~\cite{BF-book}.

The cardinality of a finite set $S$ will be denoted by $\#S$. For a prime power $q$, $\mathbb{F}_q$ denotes the finite field of order $q$ consisting of $q$ elements. 
In particular, $\mathbb{F}_2$ denotes the finite field of two elements. For a positive integer $n$, $\mathbb{F}_2^n$ is the vector space of dimension $n$ over $\mathbb{F}_2$.
The addition operation over both $\mathbb{F}_2$ and $\mathbb{F}_2^n$ will be denoted by $\oplus$. Elements of $\mathbb{F}_2^n$ are considered to be $n$-bit binary strings.

For an $n$-bit binary string $\mathbf{x}=(x_1,\ldots,x_n)$, %$\sym{wt}(\mathbf{x})$ denotes the weight of $\mathbf{x}$, i.e. 
$\sym{wt}(\mathbf{x})=\#\{i:x_i=1\}$. 
Given two strings $\mathbf{x}$ and $\mathbf{y}$ of the same length, the distance between them, denoted $d(\mathbf{x},\mathbf{y})$, is defined to be the number of
places where $\mathbf{x}$ and $\mathbf{y}$ are unequal. 
Given $\mathbf{x}=(x_1,\ldots,x_n),\mathbf{y}=(y_1,\ldots,y_n)\in \mathbb{F}_2^n$, their inner product $\langle \mathbf{x},\mathbf{y}\rangle$ is
defined to be $\langle \mathbf{x},\mathbf{y}\rangle = x_1y_1 \oplus \cdots \oplus x_ny_n$.
For an $n$-bit string $\mathbf{x}$, by $\sym{int}(\mathbf{x})$ we denote the unique integer $i\in\{0,\ldots,2^n-1\}$ whose $n$-bit binary representation is $\mathbf{x}$. 
Conversely, for $0\leq i\leq 2^n-1$, by $\sym{bin}_n(i)$ we denote the binary string given by the $n$-bit binary representation of $i$. The $n$-bit all-zero and
all-one strings will be denoted as $\mathbf{0}_n$ and $\mathbf{1}_n$ respectively. For $\mathbf{x}=(x_1,\ldots,x_n),\mathbf{y}=(y_1,\ldots,y_n)\in \mathbb{F}_2^n$, we say 
$\mathbf{x}\leq \mathbf{y}$ if $x_i\leq y_i$ for $i=1,\ldots,n$.

An $n$-variable Boolean function $f$ is a map $f:\mathbb{F}_2^n\rightarrow \mathbb{F}_2$. By $\sym{supp}(f)$ we denote the set $\{\mathbf{x}\in\mathbb{F}_2^n:f(\mathbf{x})=1\}$.
The \textit{weight} of $f$, denoted $\sym{wt}(f)$, is the size of $\sym{supp}(f)$, i.e. $\sym{wt}(f)=\#\sym{supp}(f)$. An $n$-variable function $f$ is said to be \textit{balanced}
if $\sym{wt}(f)=2^{n-1}$. An $n$-variable function $f$ is uniquely represented by a binary string $f_0\cdots f_{2^n-1}$, where for $i\in \{0,\ldots,2^n-1\}$,
$f_i=f(\sym{bin}_n(i))$. Such a string representation of $f$ is also called the \textit{truth table representation} of $f$.

\paragraph{Algebraic normal form.}
An $n$-variable function $f$ can be written as a multivariate polynomial in $\mathbb{F}_2[X_1,\ldots,X_n]/(X_1^2\oplus X_1,\ldots,X_n^2\oplus X_n)$ as follows.
Let $\mathbf{X}=(X_1,\ldots,X_n)$. Then $f(X_1,\ldots,X_n) = \bigoplus_{\bm{\alpha}\in\mathbb{F}_2^n} a_{\bm{\alpha}} \mathbf{X}^{\bm{\alpha}}$,
%\begin{eqnarray} \label{eqn-ANF}
%	f(X_1,\ldots,X_n) & = & \bigoplus_{\bm{\alpha}\in\mathbb{F}_2^n} a_{\bm{\alpha}} \mathbf{X}^{\bm{\alpha}},
%\end{eqnarray}
where $a_{\bm{\alpha}} \in \mathbb{F}_2$, and for $\bm{\alpha}=(\alpha_1,\ldots,\alpha_n)$, $\mathbf{X}^{\bm{\alpha}} = X_1^{\alpha_1}\cdots X_n^{\alpha_n}$.
This representation is called the \textit{algebraic normal form (ANF) representation} of $f$.
The algebraic degree (or simply the degree) of $f$ is defined to be $\sym{deg}(f)=\max\{\sym{wt}(\bm{\alpha}):a_{\bm{\alpha}}=1\}$. Functions of degree at most 1 are said to be
affine functions. Affine functions having $a_{\mathbf{0}_n}=0$ are said to be linear functions. It is known that if $f$ is balanced, then
$\sym{deg}(f)\leq n-1$. A balanced function $f$ with $\sym{deg}(f)=n-1$ is said to have optimal degree. 

The following equations relate the coefficients $a_{\bm{\alpha}}$ in the ANF of $f$ to the truth table representation of $f$ (see for example Pages 49 and 50 of~\cite{BF-book}).
For $\mathbf{x},\bm{\alpha}\in\mathbb{F}_2^n$,
\begin{eqnarray}\label{eqn-ANF-TT}
	f(\mathbf{x}) = \bigoplus_{\bm{\beta}\leq \mathbf{x}} a_{\bm{\beta}} & \mbox{and} &  a_{\bm{\alpha}} = \bigoplus_{\mathbf{z}\leq \bm{\alpha}} f(\mathbf{z}).
\end{eqnarray}

\paragraph{Nonlinearity and Walsh transform.}
For two $n$-variable functions $f$ and $g$, the distance between them is denoted by $d(f,g)$ and is defined to be the distance between their truth table
representations. The \textit{nonlinearity} of an $n$-variable function $f$ is denoted by $\sym{nl}(f)$ and is defined to be $\sym{nl}(f) = \min d(f,g)$, where the minimum
is over all $n$-variable affine functions $g$. 

The Walsh transform of an $n$-variable function $f$ is a map $W_f:\mathbb{F}_2^n\rightarrow \mathbb{Z}$, where for $\bm{\alpha}\in\mathbb{F}_2^n$,
$W_f(\bm{\alpha}) = \sum_{\mathbf{x}\in\mathbb{F}_2^n} (-1)^{f(\mathbf{x}) \oplus \langle \bm{\alpha}, \mathbf{x} \rangle}$. 
%\begin{eqnarray*}
%	W_f(\bm{\alpha}) & = & \sum_{\mathbf{x}\in\mathbb{F}_2^n} (-1)^{f(\mathbf{x}) \oplus \langle \bm{\alpha}, \mathbf{x} \rangle}. 
%\end{eqnarray*}
The function $f$ is balanced if and only if $W_f(\mathbf{0}_n)=0$. 
The nonlinearity of a function $f$ is given by its Walsh transform as follows:
	$\sym{nl}(f) = 2^{n-1} - \frac{1}{2}\max_{\bm{\alpha} \in \mathbb{F}_2^n} |W_f(\bm{\alpha})|$.
%\begin{eqnarray*}
%	\sym{nl}(f) & = & 2^{n-1} - \frac{1}{2}\max_{\bm{\alpha} \in \mathbb{F}_2^n} |W_f(\bm{\alpha})|.
%\end{eqnarray*}

A function $f$ such that $W_f(\bm{\alpha})=\pm 2^{n/2}$ for all $\bm{\alpha}\in\mathbb{F}_2^n$ is said to be a bent function~\cite{DBLP:journals/jct/Rothaus76}. 
Clearly such functions can exist only if 
$n$ is even. The nonlinearity of an $n$-variable bent function is $2^{n-1} - 2^{n/2-1}$ and this is the maximum nonlinearity that can be attained by $n$-variable functions.
%From Parseval's equation we have $\sum_{\bm{\alpha}\in\mathbb{F}_2^n} \left( W_f(\bm{\alpha}) \right)^2 = 2^{2n}$. 
By $\sym{LLB}(f)$ we will denote the {\em logarithm} (to base two) of the linear bias of the function $f$ which is defined in the following manner:
	$\sym{LLB}(f) = \log_2\left( 1/2 - \sym{nl}(f)/2^n \right)$.
%We would like to underline that we work with the logarithm of
%the linear bias rather than the linear bias itself. This is because for cryptographic applications, the linear bias is likely to be a small number and for small numbers
%it is more convenient to work with their logarithms than the numbers themselves.
%\begin{eqnarray}\label{eqn-LB}
%	\sym{LLB}(f) & = & \log_2\left( \frac{1}{2} - \frac{\sym{nl}(f)}{2^n} \right).
%\end{eqnarray}

For a positive integer $n$, the covering radius bound $\sym{CRB}_n$ is defined to be
$\sym{CRB}_n = 2^{n-1}-\lfloor 2^{n/2-1}\rfloor$.
%\begin{eqnarray}\label{eqn-CR}
%	\sym{CR}_n & = & 2^{n-1}-\lfloor 2^{n/2-1}\rfloor.
%\end{eqnarray}
For an $n$-variable function $f$, we have $\sym{nl}(f)\leq \sym{CRB}_n$, where equality holds for bent functions. 
Let $\sym{LCRB}_n=\log_2\left(1/2-\sym{CRB}_n/2^n\right)$.
%By $\sym{LCRB}_n$ we will denote the following quantity.
%\begin{eqnarray}\label{eqn-LCRB}
%	\sym{LCRB}_n & = & \log_2\left( \frac{1}{2} - \frac{\sym{CR}_n}{2} \right).
%\end{eqnarray}

\paragraph{Algebraic resistance.}
The \textit{algebraic immunity} of a function $f$, denoted by $\sym{AI}(f)$, is defined in the 
following manner~\cite{DBLP:conf/eurocrypt/CourtoisM03,DBLP:conf/eurocrypt/MeierPC04}:
$\sym{AI}(f)=\min_{g\neq 0} \{\sym{deg}(g): \mbox{ either } gf=0, \mbox{ or }  g(f\oplus 1)=0\}$.
%\begin{eqnarray}\label{eqn-AI}
%	\sym{AI}(f) & = & \min_{g\neq 0} \{\sym{deg}(g): \mbox{ either } gf=0, \mbox{ or }  g(f\oplus 1)=0\}.
%\end{eqnarray}
For an $n$-variable function $f$, it is known~\cite{DBLP:conf/eurocrypt/CourtoisM03} that $\sym{AI}(f)\leq \lceil n/2\rceil$.
So a function $f$ has optimal AI if $\sym{AI}(f)=\lceil n/2\rceil$.
It was proved in~\cite{DBLP:journals/tit/Didier06} that a random $n$-variable function almost surely has AI at least $\lfloor n/2-\log n\rfloor$. 
%This was more or less expected: according to Didier’s result in [F. Didier. A new upper bound on the block error probability after decoding over the erasure channel. {\em IEEE Transactions on Information Theory}  52, pp. 4496- 4503, 2006.], a random function has an AI of at least \lfloor n/2-log n\rfloor; this makes concretely that a majority of random functions have an AI of n/2 when n is even and (n-1)/2 when n is odd (but some should have an AI of (n+1)/2). Assuming that our functions behave as random ones with respect to the AI, your results are coherent, and there are chances that cases where the AI equals (n+1)/2 appear, but this may be for values of n that are too large for being reached.
%Of course, optimal AI is mathematically better, but we do not need it cryptographically (in fact, we needed an AI of around \lambda n/2 for some \lambda \leq 1 and we have better that that); the weak point of all known fast functions is their nonlinearity and this is where we need to improve.
%Maybe you can add these observations in the paper since you have the hand on it.

Algebraic immunity quantifies the resistance of a function to algebraic attacks. In practice, it is also required to provide resistance to fast algebraic 
attack (FAA)~\cite{DBLP:conf/crypto/Courtois03}.
Given an $n$-variable function $f$, let $g$ be an $n$-variable function of degree $e$ such that $gf$ has degree $d$. If for small $e$, $d$ is not too
high then the function $f$ is susceptible to an FAA. It is known~\cite{DBLP:conf/crypto/Courtois03} that for $e+d\geq n$, there exists functions $g$ and $h$ with 
$\sym{deg}(g)=e$ and $\sym{deg}(h)\leq d$ such that $gf=h$. Based on this observation, we provide the following definition.
For each $e \in \{1,\ldots,\sym{AI}(f)-1\}$, let $d\leq n-1-e$ be the maximum integer such that 
there do not exist $n$-variable functions $g$ and $h$ with $\sym{deg}(g)=e$, $\sym{deg}(h)=d$ and $gf=h$. We call the list of all such pairs $(e,d)$ 
as the \textit{FAA-profile} of $f$.

A combined measure of resistance offered by a function $f$ to both algebraic and fast algebraic attacks is defined to be 
\textit{fast algebraic immunity (FAI)}:
$$\sym{FAI}(f)\allowbreak =\allowbreak \min \left( 2\sym{AI}(f), \allowbreak 
\min_{g\neq 0}\{\sym{deg}(g)\allowbreak +\allowbreak \sym{deg}(fg): \allowbreak 1\allowbreak \leq \sym{deg}(g) \allowbreak < \allowbreak \sym{AI}(f)\}\right).$$
%\begin{eqnarray}\label{eqn-FAI}
%	\sym{FAI}(f) & = & \min \left( 2\sym{AI}(f), \min_{g\neq 0}\{\sym{deg}(g)+\sym{deg}(fg): 1\leq \sym{deg}(g) < \sym{AI}(f)\}\right).
%\end{eqnarray}
We have $\sym{FAI}(f)\allowbreak =\allowbreak \min(2\sym{AI}(f), \allowbreak \min\{e+d+1\})$, where the second minimum is taken over all pairs 
$(e,d)$ in the FAA-profile of $f$. Further, it is clear
that for any function $f$, $1+\sym{AI}(f)\leq \sym{FAI}(f)\leq 2\,\sym{AI}(f)$.

If $\sym{AI}(f)=\lceil n/2\rceil$ and for each pair $(e,d)$ in the FAA-profile of $f$, $e+d=n-1$, then $f$ is said to have perfect algebraic 
immunity (PAI)~\cite{DBLP:conf/asiacrypt/LiuZL12}. We introduce a relaxed version of the notion of optimal AI and PAI.
We say that a function $f$ has almost optimal AI if $\sym{AI}(f)\geq \lfloor n/2\rfloor$ and 
$f$ is said to have almost perfect FAA-profile if for each pair $(e,d)$ in the FAA-profile of $f$, $e+d\geq n-2$.

\begin{remark}\label{rem-AI-eff}
	There are known algorithms~\cite{DBLP:conf/eurocrypt/ArmknechtCGKMR06,DBLP:conf/fse/DidierT06,DBLP:conf/indocrypt/Didier06} for computing AI and FAI.
	The complexities of these algorithms are very high. For computing algebraic immunities we used the Boolean function
library\footnote{\url{https://doc.sagemath.org/html/en/reference/cryptography/sage/crypto/boolean_function.html\#sage.crypto.boolean_function.BooleanFunction.annihilator}} of
the SageMath software. On the computer resources available to us, it was not possible to do any computation related to algebraic resistance for functions on more than 20 variables.
\end{remark}

\paragraph{Implementation efficiency.}
The complexity of implementing a Boolean function is measured with respect to space and time. For example, a truth table representation of an $n$-variable Boolean function requires
$2^n$ bits and can be computed at a single point in $O(1)$ time (assuming that a look-up into the truth table requires constant time which need not be true if $n$
is large). More generally, we say that a Boolean function has an $(S,T)$-implementation if it can be implemented using $S$ bits/gates and can be computed using
$T$ bit operations. In an asymptotic sense, we say that an infinite family of Boolean functions has an efficient implementation if any $n$-variable function in the family 
has an $(S,T)$-implementation where both $S$ and $T$ are bounded above by polynomials in $n$. From a concrete point of view, on the other hand, we will be interested in 
the concrete details of the implementation in terms of the actual number bits required to represent the function and the actual number of basic operations required to 
compute it.

\paragraph{Vectorial functions.}
For positive integers $n$ and $m$, an $(n,m)$-vectorial Boolean function (also called an S-box) $F$ is a map $F:\mathbb{F}_2^n\rightarrow \mathbb{F}_2^m$. If $m=1$,
then we get back a Boolean function. An $(n,m)$-vectorial Boolean function $F$ can be written as $F=(f_1,\ldots,f_m)$, where each $f_i$, $i=1,\ldots,m$, is 
an $n$-variable Boolean function. The $f_i$'s are said to be the coordinate functions of $F$. For $\bm{\alpha}=(\alpha_1,\ldots,\alpha_m)\in\mathbb{F}_2^m$, let
$F_{\bm{\alpha}}=\langle \bm{\alpha},(f_1,\ldots,f_m)\rangle=\alpha_1f_1\oplus\cdots\oplus \alpha_mf_m$. Then $F_{\bm{\alpha}}$ is an $n$-variable Boolean function,
and the $F_{\bm{\alpha}}$'s are called the component functions of $F$. For $n\geq m$, an $(n,m)$-vectorial function $F$ is said to be balanced if for each 
$\bm{\beta}\in\mathbb{F}_2^m$, $\#F^{-1}(\bm{\beta})=2^{n-m}$. Equivalently, it is known that (see e.g.~\cite{BF-book}) $F$ is balanced if and only if 
all non-zero component functions of $F$ are balanced. 

Let $F$ be an $(n,m)$-vectorial Boolean function and $g$ be an $m$-variable Boolean function.
The composition $g\circ F$ is an $n$-variable Boolean function given by $(g\circ F)(X_1,\ldots,X_n) = g(F(X_1,\ldots,X_n))=g(f_1,\ldots,f_m)$. The Walsh transform
of $f\circ F$ is the following~\cite{DBLP:journals/tit/GuptaS05a}. For $\bm{\beta}\in\mathbb{F}_2^n$, 
%$W_{f\circ F}(\bm{\beta})=\frac{1}{2^m}\sum_{\bm{\alpha}\in\mathbb{F}_2^m} W_f(\bm{\alpha}) W_{F_{\bm{\alpha}}}(\bm{\beta})$.
\begin{eqnarray} \label{eqn-WT-comp}
	W_{f\circ F}(\bm{\beta}) & = & \frac{1}{2^m}\sum_{\bm{\alpha}\in\mathbb{F}_2^m} W_f(\bm{\alpha}) W_{F_{\bm{\alpha}}}(\bm{\beta}).
\end{eqnarray}
%The following simple result can be proved directly by counting pre-images and also follows from~\eqref{eqn-WT-comp}.
The following simple result follows from~\eqref{eqn-WT-comp}.
\begin{proposition}\label{prop-bal-comp}
	Let $n$ and $m$ be positive integers with $n\geq m$, and let $F$ be a balanced $(n,m)$-vectorial function. Let $f$ be an $m$-variable Boolean function.
	Then $f\circ F$ is balanced if and only if $f$ is balanced. 
\end{proposition}

\section{Relevant Previous Constructions \label{sec-prev}}
In this section, we briefly outline some previous relevant constructions.

\paragraph{Carlet-Feng (CF) functions.} Any polynomial $a(x)=a_0\oplus a_1x\oplus \cdots +a_{n-1}x^{n-1} \in \mathbb{F}_2[x]$ is uniquely determined by the
coefficient vector $\mathbf{a}=(a_{n-1},\ldots,a_0)\in \mathbb{F}_2^n$. So the elements of $\mathbb{F}_2^n$ can be considered to be polynomials in
$\mathbb{F}_2[x]$ of degree at most $n-1$. Let $\tau(x)$ be a primitive polynomial of degree $n$ over $\mathbb{F}_2$. An $n$-variable CF-function is defined
by its support which is the following set of polynomials of degrees at most $n-1$: $$\{0,1,x\bmod \tau(x),x^2\bmod \tau(x),\ldots,x^{2^{n-1}-2}\bmod \tau(x)\}.$$
It was shown in~\cite{DBLP:conf/asiacrypt/CarletF08} that such a Boolean function is balanced, has degree $n-1$ and AI $\lceil n/2\rceil$. 
(This class of functions was earlier considered in~\cite{DBLP:journals/dcc/FengLY09} for showing the tightness of bounds on the algebraic immunity of vectorial functions
and the nonlinearity was earlier studied in~\cite{DBLP:conf/wcc/BrandstatterLW05}.)
Further, it was shown in~\cite{DBLP:conf/asiacrypt/LiuZL12} that when
$n$ is one more than a power of two, such functions possess PAI. A lower bound on the nonlinearity of such functions was proved in~\cite{DBLP:conf/asiacrypt/CarletF08}.
For concrete values of $n$, the actual nonlinearities are much higher than the lower bound. Further, the nonlinearity depends
on the choice of the primitive polynomial $\tau(x)$. We computed the nonlinearities of CF functions for certain values of $n$. The primitive polynomials that we
used are given in Appendix~\ref{app-prim}.

A drawback of the CF functions is that these are not very efficient to implement. Evaluating the value of a CF function on a particular input
$a(x)$ amounts to computing $i$ such that $a(x)\equiv x^i\bmod p(x)$. This is the discrete logarithm problem in $\mathbb{F}_{2^n}$. 
A truth table implementation of CF-functions requires $O(2^n)$ bits. Using polynomial space the discrete logarithm problem can be solved in asymptotically sub-exponential time. 
As a result, CF functions are unsuitable for fast and light weight implementations.
%However, for concrete instances, the function is less slow to compute than it could seem: for particular choices of $n$, it may be possible to use the Pohlig-Hellman 
%algorithm to obtain a faster algorithm. 
%If $2^n-1$ is the product of small factors (this is the case of $n=18$ and $n=20$ for instance), it is possible to compute one output bit per cycle with 40,000 transistors, 
%as indicated in~\cite{DBLP:journals/tit/Carlet11}.

\paragraph{Hidden weight bit (HWB) functions.}
For $n\geq 1$, let $\sym{HWB}_n:\{0,1\}^n\rightarrow \{0,1\}$ be the hidden weight bit function~\cite{DBLP:journals/tc/Bryant91} defined as follows. 
For $\mathbf{x}=(x_1,\ldots,x_n)\in\mathbb{F}_2^n$,
\begin{eqnarray}\label{eqn-HWB}
	\sym{HWB}_n(\mathbf{x}) & = & x_{\sym{wt}(\mathbf{x})},
\end{eqnarray}
where we assume that $x_0=0$. The HWB functions are clearly efficiently implementable. Cryptographic properties of HWB functions were studied in~\cite{WCST}.
It was shown that the AI of $\sym{HWB}_n$ is at least $\lfloor n/3\rfloor + 1$ and for $n$ in the set $\{6,\ldots,13\}$, the actual AI is either the lower bound
or one more than the lower bound. For $n$ in the set $\{6,\ldots,13\}$, the FAA-profiles were reported in~\cite{WCST} and turned out to be significantly away from 
the profile of a PAI function. 

The nonlinearity of $\sym{HWB}_n$ was shown to be $2^{n-1}-2{n-2 \choose \lceil (n-2)/2\rceil}$. This value is quite low. So even
though HWB functions are efficiently implementable, they do not possess sufficiently high nonlinearity for cryptographic applications.
Concatenations of HWB functions have been studied in~\cite{DBLP:journals/amco/WangTS14} producing functions with higher nonlinearities than the HWB functions, but
still not high enough for use in practical systems.

Binary decision diagrams (BDD) have been used to propose attacks on stream ciphers~\cite{DBLP:conf/eurocrypt/Krause02,DBLP:conf/fse/KrauseS06}. 
A positive feature of HWB functions is that these functions have high BDD complexity~\cite{DBLP:journals/tc/Bryant91,DBLP:journals/ita/BolligLSW99,Kn09}.

\paragraph{Generalised HWB (GHWB) functions.} A generalisation of HWB functions was introduced in~\cite{DBLP:journals/tit/Carlet22} with the goal of improving their nonlinearity 
and algebraic immunity while retaining the efficiency of implementation. The concrete results for $n=13,14,15$ and $16$ presented in~\cite{DBLP:journals/tit/Carlet22} show that 
the AI of GHWB is almost optimal and is greater than the AI of HWB. There is also improvement in nonlinearity. This improvement, however, is not substantial and the obtained 
nonlinearities of GHWB functions are still not good enough for practical applications.

\paragraph{Cyclic weightwise functions.} Another generalisation of the HWB function was made in~\cite{MO24}. 
%The set $\mathbb{F}_2^n$ is partitioned into $n+1$ weight classes $W_0,\ldots,W_n$, such that vectors in $W_i$ have weight equal to $i$. 
Let $g_0,\ldots,g_n$ be $n$-variable functions. Using these $n+1$ functions,
an $n$-variable weightwise function $f$ is constructed as follows: for $\mathbf{x}\in\mathbb{F}_2^n$, $f(\mathbf{x})=g_w(\mathbf{x})$, where $w=\sym{wt}(\mathbf{x})$. 
The function $f$ is uniquely defined by the sequence of functions $(g_0,\ldots,g_n)$. Note that the function $g_w$ is applied only to strings of weight $w$. In 
particular $g_0$ is applied only to the string $\mathbf{0}_n$.
%Note that while the domain of all the $g_i$s is $\mathbb{F}_2^n$, any
%particular function $g_w$ is applied to only vectors in $\mathbb{F}_2^n$ of weight $w$. 

Since implementing $n+1$ functions may be difficult in practice, the notion of \textit{cyclic weightwise} functions was introduced in~\cite{MO24}, where the functions $g_i$'s
are defined from a single $n$-variable function $g$ as follows: $g_0=g_1=g$, and for $i\in\{2,\ldots,n\}$, $g_i$ is defined to be 
$g_i(\mathbf{x})=g(\mathbf{x} \gg\!\!> (i-1))$, where $\gg\!\!>$ is the cyclic right shift operator. The resulting function $f$ is called a cyclic weightwise function, 
which we denote as $f=\sym{CW}_n(g)$. Lower bounds on the nonlinearities of $\sym{CW}_n(g)$ was obtained in~\cite{MO24} for the case when $g$ is linear and for a
particular quadratic function $g$. For the choice of $g(x_1,\ldots,x_n)=x_1\oplus \left(\bigoplus_{i=1}^{\lfloor (n-1)/2 \rfloor}x_{2i}x_{2i+1}\right)$, actual nonlinearities, 
degrees and algebraic immunities of $\sym{CW}_n(g)$ were provided in~\cite{MO24}. These functions achieve both the highest nonlinearities and the highest algebraic
immunities among all the functions presented in~\cite{MO24}. Cyclic weightwise Boolean functions possessing properties which improve upon the
functions reported in~\cite{MO24} were described in~\cite{cryptoeprint:2024/2022}. 

\paragraph{Inverse map.} Let $\rho(x)\in \mathbb{F}_2[x]$ be an irreducible polynomial of degree $n$. Then for any nonzero polynomial $a(x)\in \mathbb{F}_2[x]$ of degree at 
most $n-1$, there is a polynomial $b(x)$ also of degree at most $n-1$ such that $a(x)b(x)\equiv 1\bmod \rho(x)$, i.e. $b(x)=a(x)^{-1}\bmod \rho(x)$. As in the case of the
CF functions, we identify polynomials in $\mathbb{F}_2[x]$ of degrees at most $n-1$ with the elements of $\mathbb{F}_2^n$. We can then define an $(n,n)$-vectorial function
$\sym{inv}:\mathbb{F}_2^n\rightarrow \mathbb{F}_2^n$ as follows: $\sym{inv}(\mathbf{0}_n)=\mathbf{0}_n$ and for any $a(x)\in \mathbb{F}_2[x]$ of degree at most $n-1$,
$\sym{inv}(a(x))=a(x)^{-1}\bmod \rho(x)$. This is the well known inverse map which was introduced to cryptography in~\cite{DBLP:conf/eurocrypt/Nyberg93}. 
A nonzero component function
of $\sym{inv}$ is an $n$-variable Boolean function. Such functions are balanced and have degrees equal to $n-1$. 
Further, it is known~\cite{DBLP:conf/eurocrypt/Nyberg93,CU57} that the nonlinearity of any non-zero component function is at least $2^{n-1}-2^{n/2}$. The AI of such a function, 
however, is not good. It was shown in~\cite{DBLP:journals/jossac/FengG16}, that the AI is equal to $\lceil 2\sqrt{n}\rceil-2$. From an implementation point of view, 
computing $a(x)^{-1}\bmod \rho(x)$ requires about $O(n^3)$ bit operations. For values of $n$ which are relevant to the nonlinear filter model, it is not possible to
make a lightweight and fast implementation of the inverse function.

%\section{Construction of $\lambda$-HWB Functions \label{sec-lambdaHWB}}
\section{Construction of Interval $\lambda$-HWB Functions \label{sec-int-lambdaHWB}}
%For $n\geq 1$, the hidden weight bit function is defined as follows (see also Appendix~\ref{sec-prev}): 
%$\sym{HWB}_n(\mathbf{0}_n)=0$ and for $\mathbf{0}_n\neq \mathbf{x}=(x_1,\ldots,x_n)\in\mathbb{F}_2^n$, $\sym{HWB}_n(\mathbf{x}) = x_{\sym{wt}(\mathbf{x})}$.
The HWB function is efficient to implement. Its major drawback, however, is its low nonlinearity. 
We provide two methods to improve the nonlinearity. The first using a post-processing and the second using a pre-processing.

\subsection{Post-processing \label{subsec-post-process}}
One possible way to improve the cryptographic
properties of the HWB function is to perform some post-processing of its output. Note that the HWB function produces a single bit of output. It is not meaningful to perform
any post-processing on a single bit. So as a first step, we consider a vectorial version of the HWB function which produces more than one bit of output. Let $r$ be the
number of bits that are to be produced. The question then is how should these $r$ bits be extracted. On an input $\mathbf{x}=(x_1,\ldots,x_n)$, the HWB function produces as output
$x_i$, where $i$ is the weight of $(x_1,\ldots,x_n)$. To extract $r$ bits, we extract a window of $r$ bits of $\mathbf{x}$ centered at $x_i$. This creates a difficulty
if indices of the window fall outside the range $\{1,\ldots,n\}$. There are two ways to tackle this situation, namely the null and the cyclic boundary conditions.
Let $\mathbf{x}=(x_1,\ldots,x_n)$ and suppose $i$ is an integer which is not in $\{1,\ldots,n\}$. Under the \textit{null boundary condition}, we define $x_i$ to be 0, while
under the \textit{cyclic boundary condition}, we define $x_i$ to be equal to $x_j$, where $j$ is the unique integer in $\{1,\ldots,n\}$ such that $i\equiv j\bmod n$.
From experimental results we find that the nonlinearities of the functions obtained using the cyclic boundary condition are more than the nonlinearities of the functions
obtained using the null boundary condition. In view of this, we do not formally introduce the construction using the null boundary condition.
%(see Remark~\ref{rem-null-vs-cyc} later). 

Given positive integers $n$ and $r$ with $r\leq n$, we define an $(n,r)$-vectorial function $\sym{HWB}_{n,r}$ as follows. 
For $\mathbf{x}\in\mathbb{F}_2^n$, let $w=\sym{wt}(\sym{x})$. 
%\begin{quote}
Let $\ell=w-\lfloor r/2\rfloor$ if $r$ is odd and let $\ell=w-r/2+1$ if $r$ is even. 
%\end{quote}
Then
\begin{eqnarray}\label{eqn-HWB-Sbox}
	\begin{array}{rcll}
		\sym{HWB}_{n,r} & = & (x_{\ell},x_{\ell+1},\ldots, x_{\ell+r-1}) & \mbox{with cyclic boundary condition}.
	\end{array}
\end{eqnarray}
Note that $\sym{HWB}_{n,1}=\sym{HWB}_{n}$. We have the following result regarding the balancedness of $\sym{HWB}_{n,r}$.
\begin{proposition}\label{prop-bal-HWB-c}
Let $n$ and $r$ be positive integers with $1\leq r\leq n$. Then $\sym{HWB}_{n,r}$ is balanced.
\end{proposition}
\begin{proof}
	Let $\bm{\beta}\in \mathbb{F}_2^r$. We count the number of preimages of $\bm{\beta}$ under $\sym{HWB}_{n,r}$. 
	For $\mathbf{x}\in \mathbb{F}_2^n$ with $w=\sym{wt}(\mathbf{x})$, suppose
	$\sym{HWB}_{n,r}(\mathbf{x})=\bm{\beta}$. Then $(x_{\ell},x_{\ell+1},\ldots, x_{\ell+r-1})=\bm{\beta}$, where $\ell=w-\lfloor r/2\rfloor$ if $r$ is 
	odd and let $\ell=w-r/2+1$ if $r$ is even. Let $k=\sym{wt}(\bm{\beta})$. Then $\#\{i\in \{1,\ldots,n\}\setminus \{\ell,\ldots,\ell+r-1\}: x_i=1\}=w-k$. So
	the number of $\mathbf{x}$'s such that $\sym{wt}(\mathbf{x})=w$ and $(x_{\ell},x_{\ell+1},\ldots, x_{\ell+r-1})=\bm{\beta}$ is equal to 
	${n-r\choose w-k}$. Consequently, the number of preimages of $\bm{\beta}$ under $\sym{HWB}_{n,r}$ is $\sum_{w=0}^n{n-r\choose w-k}=2^{n-r}$, since $n-k\geq n-r$. 
\end{proof}

Let $\lambda$ be an $r$-variable Boolean function. We define an $n$-variable Boolean function $\lambda\mbox{-}\sym{HWB}_{n,r}$ in the following manner.
\begin{eqnarray}\label{eqn-lambda-HWB}
	\lambda\mbox{-}\sym{HWB}_{n,r} & = & \lambda \circ \sym{HWB}_{n,r}.
\end{eqnarray}
So for $\mathbf{x}\in\mathbb{F}_2^n$, $\lambda\mbox{-}\sym{HWB}_{n,r}(\mathbf{x})=\lambda(\sym{HWB}_{n,r}(\mathbf{x}))$.

%> f_0(x1,...,x8) = g(x8,x1,...,x7) 			g(x2,...,x8,x1)
%> f_1(x1,...,x8) = g(x1,...,x8)			g(x1,...,x8)
%> f_2(x1,...,x8) = g(x2,...,x8,x1)			g(x8,x1,...,x7)
%> f_3(x1,...,x8) = g(x3,...,x8,x1,x2) and so on.	g(x7,x8,x1,...,x6)
%>
%> Since f=\lambda-HWB, we have
%>
%> f_0(x1,...,x8) = lambda(0,0,0)			lambda(0,0,0)
%> f_1(x1,...,x8) = lambda(x8,x1,x2)			lambda(x8,x1,x2)
%> f_2(x1,...,x8) = lambda(x1,x2,x3)			lambda(x1,x2,x3)
%> f_3(x1,...,x8) = lambda(x2,x3,x4) and so on.		lambda(x2,x3,x4)
%Take g(x1,…,x8)=\lambda(x8,x1,x2). You have g(x2,...,x8,x1)=\lambda(x1,x2,x3), g(x3,...,x8,x1,x2) =\lambda(x2,x3,x4) so f1,f2,f3 coincide and f0 coincide since the input is 0.

\begin{proposition}\label{prop-comp-bal-hwb-lambda}
Let $\lambda$ be an $r$-variable Boolean function. Then $\lambda\mbox{-}\sym{HWB}_{n,r}$ is balanced if and only if $\lambda$ is balanced. 
\end{proposition}
\begin{proof}
	Proposition~\ref{prop-bal-HWB-c} shows that $\sym{HWB}_{n,r}$ is a balanced $(n,r)$-vectorial function. From Proposition~\ref{prop-bal-comp} we have
	that the composition of a balanced $(n,r)$-vectorial function and an $r$-variable Boolean function $\lambda$ is balanced if and only if $\lambda$ is balanced. 
\end{proof}

Let $\pi_1,\ldots,\pi_n$ be permutations of $\{1,\ldots,n\}$ and for $i=1,\ldots,n$, let $P_i:\mathbb{F}_2^n\rightarrow\mathbb{F}_2^n$ be defined as
$P_i(x_1,\ldots,x_n)=(x_{\pi(1)},\ldots,x_{\pi(n)})$. Let $g$ be an $n$-variable Boolean function and $f$ be another $n$-variable Boolean function defined
using $g$ and $P_1,\ldots,P_n$ in the following manner: $f(\mathbf{x})=g(P_w(\mathbf{x}))$, where $w=\sym{wt}(\mathbf{x})$. Proposition~4 of~\cite{MO24} shows that 
$f$ is balanced if and only if $g$ is balanced.
Proposition~\ref{prop-comp-bal-hwb-lambda} can be seen as a corollary of Proposition~4 of~\cite{MO24}. 
On the other hand, Proposition~4 of~\cite{MO24} itself can be seen as a corollary of Proposition~\ref{prop-bal-comp} in the following manner. Given $P_1,\ldots,P_n$,
define a bijection $S:\mathbb{F}_2^n\rightarrow\mathbb{F}_2^n$ by $S(\mathbf{x})=P_w(\mathbf{x})$, where $w=\sym{wt}(\mathbf{x})$. Then $f=g\circ S$, and
by Proposition~\ref{prop-bal-comp}, $f$ is balanced if and only if $g$ is balanced.

\begin{proposition}\label{prop-complement}
	For any $r$-variable function $\lambda$, $(1\oplus\lambda)\mbox{-}\sym{HWB}_{n,r}=1\oplus \lambda\mbox{-}\sym{HWB}_{n,r}$.
	More generally, for any invertible affine transformation $A:\mathbb{F}_2^r\rightarrow\mathbb{F}_2^r$, 
	$\sym{nl}(\lambda\circ A\circ \sym{HWB}_{n,r}) = \sym{nl}(\lambda\circ\sym{HWB}_{n,r})$.
\end{proposition}

%\begin{remark}\label{rem-null1}
%Since $\sym{HWB}_{n,r,\sym{n}}$ is not balanced, the balancedness of $\lambda\mbox{-}\sym{HWB}_{n,r,\sym{n}}$ does not reduce to the balancedness
%of $\lambda$. It is, however, possible for $\lambda\mbox{-}\sym{HWB}_{n,r,\sym{n}}$ to be balanced. We have experimented extensively with 
%$\lambda\mbox{-}\sym{HWB}_{n,r,\sym{n}}$ to create balanced functions. The best nonlinearities that are obtained with the null boundary condition
%	are however, less than the best nonlinearities obtained using the cyclic boundary condition. (See Remark~\ref{rem-null-vs-cyc} below for further
%	discussion.) In view of this observation, we focus only on the cyclic boundary condition.
%\end{remark}

%\begin{remark}\label{rem-req1req2}
%	For $r=1$, $\lambda\mbox{-}\sym{HWB}_{n,r}=\sym{HWB}_{n}$. For $r=2$, the only balanced functions are affine functions and hence,
%	for $r=2$, if $\lambda\mbox{-}\sym{HWB}_{n,r}$ is balanced, then $\sym{nl}(\lambda\mbox{-}\sym{HWB}_{n,r})=\sym{nl}(\sym{HWB}_{n})$.
%	So in the following, we focus on $r\geq 3$.
%\end{remark}

The nonlinearity of $\lambda\mbox{-}\sym{HWB}_{n,r}$ is determined by the Walsh transform of $\lambda\mbox{-}\sym{HWB}_{n,r}$. In principle,
using~\eqref{eqn-WT-comp}, the Walsh transform of $\lambda\mbox{-}\sym{HWB}_{n,r}$ can be determined from the Walsh transforms of $\lambda$ and
$\sym{HWB}_{n,r}$. So in principle, using~\eqref{eqn-WT-comp}, the nonlinearity of $\lambda\mbox{-}\sym{HWB}_{n,r}$ can be determined from the
Walsh transforms of $\lambda$ and $\sym{HWB}_{n,r}$. The form of~\eqref{eqn-WT-comp}, however, does not provide any easy method to identify conditions
on the Walsh transform of $\lambda$ such that the nonlinearity of $\lambda\mbox{-}\sym{HWB}_{n,r}$ is high. 
%Our goal is to choose $\lambda$ such that $\lambda\mbox{-}\sym{HWB}_{n,r}$ has high nonlinearity. As discussed above, the expression for the 
%Walsh transform of $\lambda\mbox{-}\sym{HWB}_{n,r}$ given by~\eqref{eqn-WT-comp} does not provide any guidance. Further, as discussed in Remark~\ref{rem-nl-cyc-wtwise}
%the analysis in~\cite{MO24} provides loose lower bounds for some very special choices of $\lambda$. 
Faced with this scenario, we decided to search for 
choices of $\lambda$ to determine the set of $\lambda$'s having the highest possible nonlinearity. Since we are interested in balanced functions, using
Proposition~\ref{prop-comp-bal-hwb-lambda}, we focused only on balanced $\lambda$'s. 
Algorithm~\ref{algo-srch} describes our search strategy. It takes as input $n$, $r$ and a list $\mathcal{S}$ of $r$-variable balanced functions
and produces as output a set of 
functions $\lambda$ such that the corresponding $\lambda\mbox{-}\sym{HWB}_{n,r}$ function has algebraic degree $n-1$ and as such, has maximal nonlinearity among all 
visited functions.

\begin{algorithm}[h] 
\SetAlgorithmName{Algorithm}{}
\DontPrintSemicolon
	\KwInput{$n$, $r$ and $\mathcal{S}$, where $\mathcal{S}$ is a subset of the set of all balanced $r$-variable functions}
	\KwOutput{A list $\mathcal{L}$ of $r$-variable functions such that for any $\lambda\in \mathcal{L}$, $\lambda\mbox{-}\sym{HWB}_{n,r}$ is balanced,
	has degree $n-1$ and $\lambda \in \argmax_{\mu\in\mathcal{S}}\sym{nl}(\mu\mbox{-}\sym{HWB}_{n,r})$}

	$\sym{maxnl} \leftarrow 0$; $\mathcal{L}\leftarrow \emptyset$ \\
	\For{$\lambda\in \mathcal{S}$} {
		let $f=\lambda\mbox{-}\sym{HWB}_{n,r}$ \\
		compute $\sym{nl}(f)$ and $\sym{deg}(f)$ \\
		\If{$\sym{deg}(f) = n-1$ and $\sym{maxnl} < \sym{nl}(f)$} {
			$\sym{maxnl} \leftarrow \sym{nl}(f)$; $\mathcal{L}\leftarrow \{\lambda\}$
		}
		\Else {
			\If{$\sym{deg}(f) = n-1$ and $\sym{maxnl}=\sym{nl}(f)$} {
				$\mathcal{L}\leftarrow \mathcal{L} \cup \{\lambda\}$	
			}
		}
	}
	return $\mathcal{L}$
	\caption{The search procedure for $\lambda\mbox{-}\sym{HWB}_{n,r}$. \label{algo-srch}}	
\end{algorithm}	

\begin{proposition}\label{prop-bal-maxdeg}
	For positive integers $n$ and $r$ with $1\leq r\leq n$ and $\mathcal{S}$ a subset of balanced $r$-variable functions, let $\mathcal{L}$ be
	returned by Algorithm~\ref{algo-srch} on input $n$, $r$ and $\mathcal{S}$. Then for any $\lambda \in \mathcal{L}$,
	$\lambda\mbox{-}\sym{HWB}_{n,r}$ is a balanced $n$-variable function having degree $n-1$. The time taken by Algorithm~\ref{algo-srch}
	is $O(\#\mathcal{S}\, n2^n)$.
\end{proposition}
\begin{proof}
	Suppose $\mathcal{L}$ is the output of Algorithm~\ref{algo-srch}. From Proposition~\ref{prop-comp-bal-hwb-lambda} it follows that any $\lambda \in \mathcal{L}$ is balanced. 
	From the algorithm, it directly follows that the degree is $n-1$. 

For each $\lambda$ in $\mathcal{S}$, the algorithm constructs the $n$-variable function $\lambda\mbox{-}\sym{HWB}_{n,r}$ and computes its nonlinearity and degree. 
So the time for each $\lambda$ is $O(n2^n)$, and the total time is $O(\#\mathcal{S}\, n2^n)$. 
\end{proof}

If $\mathcal{S}$ is the set of all balanced $r$-variable functions, then the time required by Algorithm~\ref{algo-srch} is $O({2^r\choose 2^{r-1}} n2^n)$.
For $r=2,3$ and $4$, and for $n=13,\ldots,20$, we have run Algorithm~\ref{algo-srch} with $\mathcal{S}$ to be the set of all $r$-variable balanced Boolean functions. 
(Note that for $r=2$ the only balanced functions are the non-constant affine functions.) A summary of our observations of these executions of Algorithm~\ref{algo-srch}
are as follows.
%\begin{enumerate}
\begin{compactenum}
	\item For $n=13,\ldots,20$ and $r=2$, for the $\lambda$'s produced by Algorithm~\ref{algo-srch}, the nonlinearities of $\lambda\mbox{-}\sym{HWB}_{n,2}$ are equal 
		to the nonlinearities of the corresponding $\sym{HWB}_n$. This though is not true in general. For example, for $n=8$, taking $\lambda(X_1,X_2)=X_1\oplus X_2$, the
		nonlinearity of $\lambda\mbox{-}\sym{HWB}_{8,2}$ is 92, while the nonlinearity of $\sym{HWB}_8$ is 88.
	\item For a fixed value of $n$, the nonlinearity of $\lambda\mbox{-}\sym{HWB}_{n,r}$ with $\lambda$ produced by Algorithm~\ref{algo-srch} increases with the value
		of $r$.
\end{compactenum}
%\end{enumerate}
For $r=5$, the number of balanced $r$-variable functions is equal to ${32\choose 16}\approx 2^{29.163}$. So if in Algorithm~\ref{algo-srch} we put $\mathcal{S}$
to be the set of all 5-variable balanced functions, then the time taken will be proportional to $n2^{n+29.163}$. On the computing resources available to use, for $n=13$
this computation is barely feasible while it is out of our reach for $n=20$. Accordingly, we decided to take $\mathcal{S}$ to be a proper subset of 5-variable balanced
functions. The first condition that we imposed is to consider only functions having degree 4. This, however, does not significantly reduce the size of $\mathcal{S}$. 
Next we imposed the condition that along with degree 4, the functions should have nonlinearity 12, which is the maximum possible nonlinearity among all 5-variable 
balanced functions. This condition is motivated by our finding that for $r=3$ and $r=4$, the $\lambda$'s which are returned by Algorithm~\ref{algo-srch} have the
maximum possible nonlinearity among all balanced $r$-variable functions. 
The number of 5-variable functions having degree 4 and nonlinearity 12 is
$1666560\approx 2^{20.668}$. With $\#\mathcal{S}=1666560$, it becomes feasible to run Algorithm~\ref{algo-srch} for $n=13,\ldots,20$ on our computers. 
The nonlinearities that are obtained are higher than the nonlinearities obtained for $r=2,3$ and $4$.
The following proposition states the results that we obtained. 

\begin{proposition}\label{prop-lambda-hwb-req5}
Let $r=5$. For $n=13,\ldots,20$, the maximum nonlinearities, along with the corresponding $\lambda$'s and $1\oplus \lambda$'s, achieved by balanced 
	$\lambda\mbox{-}\sym{HWB}_{n,r}$ functions having degree $n-1$, where $\lambda$ runs over all $5$-variable balanced functions having degree 4
	and nonlinearity 12, are as follows.
	\begin{itemize}
		\item $n=13$, $\sym{nl}(\lambda\mbox{-}\sym{HWB}_{n,r})=3780$, where $\lambda,1\oplus\lambda \in \{\lambda_{5,1},\lambda_{5,2}\}$. 
		\item $n=14$, $\sym{nl}(\lambda\mbox{-}\sym{HWB}_{n,r})=7572$, where $\lambda,1\oplus\lambda \in \{\lambda_{5,3},\lambda_{5,4}\}$. 
		\item $n=15$, $\sym{nl}(\lambda\mbox{-}\sym{HWB}_{n,r})=15236$, where $\lambda,1\oplus\lambda \in \{\lambda_{5,1},\lambda_{5,2}\}$. 
		\item $n=16$, $\sym{nl}(\lambda\mbox{-}\sym{HWB}_{n,r})=30526$, where $\lambda,1\oplus\lambda \in \{\lambda_{5,5},\lambda_{5,6}\}$. 
		\item $n=17$, $\sym{nl}(\lambda\mbox{-}\sym{HWB}_{n,r})=61284$, where $\lambda,1\oplus\lambda \in \{\lambda_{5,1},\lambda_{5,2}\}$. 
		\item $n=18$, $\sym{nl}(\lambda\mbox{-}\sym{HWB}_{n,r})=122758$, where $\lambda,1\oplus\lambda \in \{\lambda_{5,7},\lambda_{5,8}\}$. 
		\item $n=19$, $\sym{nl}(\lambda\mbox{-}\sym{HWB}_{n,r})=246368$, where $\lambda,1\oplus\lambda \in \{\lambda_{5,9},\lambda_{5,10}\}$. 
		\item $n=20$, $\sym{nl}(\lambda\mbox{-}\sym{HWB}_{n,r})=493476$, where $\lambda,1\oplus\lambda \in \{\lambda_{5,11},\lambda_{5,12}\}$. 
	\end{itemize}
	In the above, $\lambda_{5,i}$, $i=1,\ldots,12$, given by their 32-bit string representations are the following. (The ANFs of these functions
	are given in Appendix~\ref{app-ANF}.)
	{\scriptsize
	\begin{eqnarray*} 
		%13,15,17: \lambda_{5,1},\lambda_{5,2}; 14: \lambda_{5,3},\lambda_{5,4}; 16: \lambda_{5,5},\lambda_{5,6}; 
		%18: \lambda_{5,7},\lambda_{5,8}; 19: \lambda_{5,9},\lambda_{5,10}; 20: \lambda_{5,11},\lambda_{5,12}
		\lambda_{5,1} = 10111111010100010001101000001110, %1884850941
		& & \lambda_{5,2} = 10101000011010110100111001001110 \\ %1920128533
		\lambda_{5,3} = 10010011011000111011010111010000, %195937993
		& & \lambda_{5,4} = 10000100110100111010100111110100 \\ %798346017
		\lambda_{5,5} = 10101011011010110001101100011000, %416863957
		& & \lambda_{5,6} = 10000101111110111000101000001110 \\ %1884413857
		\lambda_{5,7} = 11100001010111110000101001001110, %1917909639
		& & \lambda_{5,8} = 10101011001100100001111000011110 \\ %2021149909
		\lambda_{5,9} = 10001001010111110010110011101000, %389347985
		& & \lambda_{5,10} = 10101011100100111011010100000110 \\ %1622002133
		\lambda_{5,11} = 01100010101011111110000111000100, %596112710
		& & \lambda_{5,12} = 01100000110010110101011101001110 \\ %1927992070
	\end{eqnarray*}
	}
\end{proposition}

%We computed the algebraic immunities of the functions given by Proposition~\ref{prop-lambda-hwb-req5}. The results are given in the following proposition.
%\begin{proposition}\label{prop-lambda-hwb-AI}
%For $r=5$ and $n=13,\ldots,18$, the algebraic immunities of the functions described in Proposition~\ref{prop-lambda-hwb-req5} are the following.
%	%\begin{enumerate}
%	\begin{compactenum}
%		\item $\sym{AI}(\lambda_{5,1}\mbox{-}\sym{HWB}_{13,5}) = \sym{AI}(\lambda_{5,2}\mbox{-}\sym{HWB}_{13,5}) = 6$.
%		\item $\sym{AI}(\lambda_{5,3}\mbox{-}\sym{HWB}_{14,5}) = \sym{AI}(\lambda_{5,4}\mbox{-}\sym{HWB}_{14,5}) = 7$.
%		\item $\sym{AI}(\lambda_{5,1}\mbox{-}\sym{HWB}_{15,5}) = \sym{AI}(\lambda_{5,2}\mbox{-}\sym{HWB}_{15,5}) = 7$.
%		\item $\sym{AI}(\lambda_{5,5}\mbox{-}\sym{HWB}_{16,5}) = \sym{AI}(\lambda_{5,6}\mbox{-}\sym{HWB}_{16,5}) = 7$.
%		\item $\sym{AI}(\lambda_{5,1}\mbox{-}\sym{HWB}_{17,5}) = \sym{AI}(\lambda_{5,2}\mbox{-}\sym{HWB}_{17,5}) = 8$.
%		\item $\sym{AI}(\lambda_{5,7}\mbox{-}\sym{HWB}_{18,5}) = \sym{AI}(\lambda_{5,8}\mbox{-}\sym{HWB}_{18,5}) = 8$.
%	\end{compactenum}
%	%\end{enumerate}
%\end{proposition}

\paragraph{Efficiency of computing $\lambda\mbox{-}\sym{HWB}_{n,5}$.} The requirement is to compute the weight of one $n$-bit string
and to compute the output of a $5$-variable function. For $n\leq 20$, this is very efficient to do in both hardware and software. 

\paragraph{Relation to Cyclic Weightwise Functions.} Let $\ell_1=1-\lfloor r/2\rfloor$ if $r$ is odd and let $\ell_1=1-r/2+1$ if $r$ is even.
Define an $n$-variable function $g$, where for $(x_1,\ldots,x_n)\in \mathbb{F}_2^n$, $g(x_1,\ldots,x_n)=\lambda(x_\ell,x_{\ell+1},\ldots,x_{\ell+r-1})$ with cyclic boundary
condition. Let $g_0,g_1,\ldots,g_n$ be $n$-variable functions where $g_0=g_1=g$ and for $i\in \{2,\ldots,n\}$, 
$g_i(x_1,\ldots,x_n)=g((x_1,\ldots,x_n) <\!\!\ll (i-1))$, where $<\!\!\ll$ is the cyclic left shift operator.
Then $\lambda\mbox{-}\sym{HWB}_{n,r}$ is a weightwise function defined by the sequence of functions $(g_0,g_1,\allowbreak \ldots,\allowbreak g_n)$. Note that the notion of cyclic
weightwise functions is defined using right cyclic shifts, whereas $\lambda\mbox{-}\sym{HWB}_{n,r}$ is obtained from $g$ using left cyclic 
shifts\footnote{We were unaware of 
the paper~\cite{MO24} when we obtained the function $\lambda\mbox{-}\sym{HWB}_{n,r}$. It is only later that we realised that $\lambda\mbox{-}\sym{HWB}_{n,r}$ is a special case 
of (left) cyclic weightwise functions.}. 

\subsection{Pre-processing \label{subsec-pre-process}}
The function $\lambda\mbox{-}\sym{HWB}_{n,r}$ improves the properties of the HWB function by first extending the HWB function to a vectorial function
and then applying $\lambda$ to the output of the vectorial function. This constitutes a post-processing of the output of the HWB vectorial function. 
%The functions that are obtained using this approach provide a different trade-off from the functions reported in Table~7 of~\cite{MO24}, while the nonlinearities
%are lower, the degrees and the algebraic immunities are the same or higher. 

To further improve the nonlinearity, we consider a pre-processing of the input to $\lambda\mbox{-}\sym{HWB}_{n,r}$. In more details, we construct a nonlinear bijection
$\phi:\mathbb{F}_2^n\rightarrow \mathbb{F}_2^n$, so that before applying $\lambda\mbox{-}\sym{HWB}_{n,r}$ to an input $\mathbf{x}\in \mathbb{F}_2^n$, we
first apply $\phi$ to $\mathbf{x}$ to obtain $\mathbf{y}$ and then apply $\lambda\mbox{-}\sym{HWB}_{n,r}$ to $\mathbf{y}$. 

The bijection $\phi$ combines integer and binary
field arithmetic. Given $\mathbf{x}\in \mathbb{F}_2^n$, $\phi$ does the following: 
changes the representation of $\mathbf{x}$ to an element of $\mathbb{Z}_{2^n}$ and applies a bijection $\mathcal{B}$; 
changes back the representation to $\mathbb{F}_2^n$ and reverses the string (which is a linear operation over $\mathbb{F}_2^n$);
again changes the representation to $\mathbb{Z}_{2^n}$ and applies $\mathcal{B}$; changes back representation to $\mathbb{F}_2^n$ and produces the output. 
Note that changing representations from $\mathbb{F}_2^n$ to $\mathbb{Z}_{2^n}$ and vice versa is simply a matter of considering the input to be either a binary string
or a non-negative integer, and has no cost. The bijection $\mathcal{B}$ considers $\mathbb{Z}_{2^n}$ to be partitioned into intervals; determines the interval to which
the integer representation $i$ of $\mathbf{x}$ belongs, applies a simple permutation of the interval to $i$ to obtain $j$ which is then returned. 

The above strategy for constructing $\phi$ was determined after a great deal of experimentation with combining simple integer and binary field arithmetic. 
For example, the bijection $\mathcal{B}$ is applied twice; our experiments show that applying $\mathcal{B}$ twice rather than once leads to noticeable increase
in nonlinearity, but futher application of $\mathcal{B}$ does not provide significant gain in nonlinearity.
We also considered various other approaches, but the nonlinearities achieved by such constructions were not found to be sufficiently high. 
The approach that we describe achieves good nonlinearity as well as good algebraic resistance as we report below.

Let $\mathbb{Z}_{2^n}$ be the set of integers modulo $2^n$. We construct $\phi$ by mixing simple and fast operations over $\mathbb{Z}_{2^n}$ and $\mathbb F_{2^n}$
(conversions between the representations $\mathbb{F}_{2^n}$ and $\mathbb Z_{2^n}$ are done using the functions $\sym{int}(\mathbf{x})$ and $\sym{bin}_n(i)$,
as described in Section~\ref{sec-prelim}.)
The fact that each of the structures $\mathbb{Z}_{2^n}$ and $\mathbb{F}_{2^n}$ is complex with respect to the other is used in the so-called ARX cryptosystems. 

The core of our construction of $\phi$ is based on the idea of partitioning $\mathbb{Z}_{2^n}$ into intervals. We first describe this partitioning strategy.
\begin{quote}
	\textit{Partition of $\mathbb{Z}_{2^n}$}: Let $n\geq 2$ and $s<n$ be a positive integer.  
	Let $0\leq w_0,\ldots,w_{2^s-1}\leq 2^n-1$ be integers such that $w_{k+1}=w_k+2^{n-s}\bmod 2^n$. 
For $0\leq k\leq 2^s-1$, let $I_k=\{w_k,w_k+1,\ldots,w_k+2^{n-s}-1\}$ where the elements of the set $I_k$ are computed modulo $2^n$. 
\end{quote}

\begin{proposition}\label{prop-part}
	The collection of sets $\{I_k\}$ with $k=0,\ldots,2^s-1$ forms a partition of $\mathbb{Z}_{2^n}$.
\end{proposition}
\begin{proof}
	Note that the number of $I_k$'s is $2^{s}$, and each $I_k$ is a subset of $\mathbb{Z}_{2^n}$ containing $2^{n-s}$ elements. So to show the result it is sufficient 
	to show that for $0\leq k<\ell\leq 2^s-1$, $I_k$ and $I_\ell$ are disjoint. From the definition of the 
	$w_k$'s, we have $w_\ell=w_k+(\ell-k)2^{n-s}\bmod 2^n$. Suppose that $I_k$ and $I_\ell$
	have a non-empty intersection. Then there are integers $a$ and $b$ with $0\leq a,b \leq 2^{n-s}-1$ such that $w_k+a\equiv w_\ell+b\bmod 2^n$, i.e.
	$(\ell-k)2^{n-s}+(b-a)\equiv 0\bmod 2^n$. 
	Note that $1\leq \ell-k\leq 2^s-1$ and so $2^{n-s}\leq (\ell-k)2^{n-s}\leq 2^n-2^{n-s}$. Further, $-2^{n-s}+1\leq b-a\leq 2^{n-s}-1$.
	So $1\leq (\ell-k)2^{n-s}+(b-a)\leq 2^n-1$. Consequently, $(\ell-k)2^{n-s}+(b-a)\not\equiv 0\bmod 2^n$, which is a contradiction.
\end{proof}

\begin{proposition}\label{prop-uni-I}
	For $n\geq 2$, $w_0\in \mathbb{Z}_{2^n}$ and positive integer $s<n$, define $\mathcal{I}_{n,w_0,s}:\mathbb{Z}_{2^n} \rightarrow \mathbb{Z}_{2^s}$ as follows.
\begin{eqnarray}\label{eqn-uni-I}
\mathcal{I}_{n,w_0,s}(i)
	& = & \left\{
	\begin{array}{ll}
		\left\lfloor \frac{i-w_0}{2^{n-s}} \right\rfloor & \mbox{if } i \geq w_0, \\ \\
		\left\lfloor \frac{i+2^n-w_0}{2^{n-s}} \right\rfloor & \mbox{if } i < w_0.
	\end{array}
	\right.
\end{eqnarray}
	Let $k=\mathcal{I}_{n,w_0,s}(i)$. Then $w_k=w_0+k 2^{n-s}\bmod 2^n$ and $k$ is the unique integer such that $i$ is in $I_k=\{w_k,w_k+1,\ldots,w_k+2^{n-s}-1\}$. 
\end{proposition}

Using the collection of intervals $\{I_k\}$, we define a bijection $\mathcal{B}$ of $\mathbb{Z}_{2^n}$. The idea is the following. Let $i\in \mathbb{Z}_{2^n}$. Then $i$ is in
one of the intervals $I_k$, and from $i$, the value of $k$ can be found using Proposition~\ref{prop-uni-I}. Suppose then that $i=w_k+a$, for some $a\in \mathbb{Z}_{2^{n-s}}$. 
Let $b=(2k+1)a\bmod 2^{n-s}$. Since $2k+1$ is odd, the map $a\mapsto (2k+1)a\bmod 2^{n-s}$ is a bijection of $\mathbb{Z}_{2^{n-s}}$. So $b\in \mathbb{Z}_{2^{n-s}}$. 
Let $j=w_k+b$. We set $\mathcal{B}(i)$ to be equal to $j$. In the following result we provide a more formal description of the bijection $\mathcal{B}$.
\begin{proposition}\label{prop-bij}
	For $n\geq 2$, positive integer $s<n$ and $w_0\in \mathbb{Z}_{2^n}$, define $\mathcal{B}_{n,w_0,s}:\mathbb{Z}_{2^n} \rightarrow \mathbb{Z}_{2^n}$ as follows.
	For $i\in \mathbb{Z}_{2^n}$, the value of $\mathcal{B}_{n,w_0,s}(i)$ is determined by the following sequence of steps. 
	\begin{tabbing}
		\ \ \= 3.\ \= \ \ \ \ \kill
		\> 1. \> $k\leftarrow \mathcal{I}_{n,w_0,s}(i)$; \\
		\> 2. \> $w_k\leftarrow w_0+k 2^{n-s}\bmod 2^n$; \\
		\> 3. \> $a\leftarrow (i-w_k)\bmod 2^n$; \\
		\> 4. \> $b\leftarrow a(2k+1)\bmod 2^{n-s}$; \\
		\> 5. \> $j\leftarrow b+w_k\bmod 2^n$; \\
		\> 6. \> set $\mathcal{B}_{n,w_0,s}(i)$ to be equal to $j$.
	\end{tabbing}
	The map $\mathcal{B}_{n,w_0,s}$ defined above is a bijection.
\end{proposition}
\begin{proof}
	From Proposition~\ref{prop-uni-I}, $k$ is the unique integer such that $i$ is in $I_k$. Since $w_0$ is given, $w_k$ is uniquely determined by $k$ and
	hence $w_k$ is uniquely determined by $i$. So $a=i-w_k\bmod 2^n$ is an element of $\mathbb{Z}_{2^{n-s}}$, which is uniquely determined by $i$.
	Since $2k+1$ is odd, the map $a\mapsto a(2k+1)\bmod 2^{n-s}$ is a bijection from $\mathbb{Z}_{2^{n-s}}$ to itself. 
	So $b$ is in $\mathbb{Z}_{2^{n-s}}$ and is uniquely determined by $a$. Since $j=b+w_k\bmod 2^n$, $b$ is uniquely determined by $a$, $a$ itself is uniquely 
	determined by $i$, and $w_k$ is uniquely determined by $i$, it follows that $j$ is also uniquely determined by $i$. 
	This shows that $\mathcal{B}_{n,w_0,s}$ is an injection and hence a bijection.
\end{proof}
%        if (intx >= w0) k = (intx-w0) >> (n-s); else k = (intx+(1<<n)-w0) >> (n-s);
%        wk = (w0 + k * (1<<(n-s))) % (1<<n);
%        intx = (intx-wk) % (1<<n);
%        intx = (intx*alphaLst[k]) % (1<<(n-s)); // intx = reverseInt(intx,n-s); intx = (intx*alphaLst[k]) % (1<<(n-s));
%        k = (k*mults)%(1<<s); // k = reverseInt(k,s); k = (k*mults)%(1<<s);
%        wk = (w0 + k * (1<<(n-s))) % (1<<n);
%        intx = (intx + wk) % (1<<n);
%
%        intx = reverseInt(intx,n);
%
%        if (intx >= w0) k = (intx-w0) >> (n-s); else k = (intx+(1<<n)-w0) >> (n-s);
%        wk = (w0 + k * (1<<(n-s))) % (1<<n);
%        intx = (intx-wk) % (1<<n);
%        intx = (intx*alphaLst[k]) % (1<<(n-s));
%        k = (k*mults)%(1<<s);
%        wk = (w0 + k * (1<<(n-s))) % (1<<n);
%        intx = (intx + wk) % (1<<n);

Given $\mathbf{x}=(x_1,x_2,\ldots,x_{n-1},x_n)\in\mathbb{F}_2^n$, let $\sym{reverse}(\mathbf{x})$ denote the string 
$(x_n,\allowbreak x_{n-1},\allowbreak \ldots,\allowbreak x_2,\allowbreak x_1)$, i.e.  $\sym{reverse}(\mathbf{x})$ reverses the string $\mathbf{x}$. 
Using $\mathcal{B}$ and $\sym{reverse}$, we define a bijection $\phi$ from $\mathbb{F}_{2}^n$ to itself. The idea is the following. Given $\mathbf{x}\in\mathbb{F}_2^n$,
change the representation to $i\in \mathbb{Z}_{2^n}$. Let $j=\mathcal{B}(i)$.
Change the representation of $j$ from $\mathbb{Z}_{2^n}$ to $\mathbb{F}_{2^n}$, use $\sym{reverse}$, and then change the representation back to $\mathbb{Z}_{2^n}$. 
Apply $\mathcal{B}$ once again and change the representation to $\mathbb{F}_{2^n}$ and produce as the output of $\phi$. The description is made precise in the
following result.
\begin{proposition}\label{prop-phi}
	Given $n\geq 2$, positive integer $s<n$ and $w_0\in \mathbb{Z}_{2^n}$, define a map $\phi_{n,w_0,s}:\mathbb{F}_2^n\rightarrow\mathbb{F}_2^n$ as follows.
	For $\mathbf{x}\in \mathbb{F}_2^n$, the following defines $\phi_{n,w_0,s}(\mathbf{x})$. 
\begin{tabbing}
	\ \ \= \ \ \ \ \kill
	\> $i\leftarrow \sym{int}(\mathbf{x})$; $j\leftarrow \mathcal{B}_{n,w_0,s}(i)$; $\mathbf{y}\leftarrow \sym{bin}_n(j)$; \\
	\> $\mathbf{w}\leftarrow \sym{reverse}(\mathbf{y})$; \\
	\> $i\leftarrow \sym{int}(\mathbf{w})$; $j\leftarrow \mathcal{B}_{n,w_0,s}(i)$; $\mathbf{z}\leftarrow \sym{bin}_n(j)$; \\
	\> set $\phi_{n,w_0,s}(\mathbf{x})$ to be equal to $\mathbf{z}$.
\end{tabbing}
The map $\phi_{n,w_0,s}$ described above is a bijection.
\end{proposition}
Using $\phi_{n,w_0,s}$ and $\lambda\mbox{-}\sym{HWB}_{n,r}$ we define a Boolean function $\sym{IntHWB}_{n,w_0,s,\lambda}:\mathbb{F}_2^n\rightarrow \mathbb{F}_2^n$
as follows. 
\begin{eqnarray}\label{eqn-int-lam-hwb}
	\sym{IntHWB}_{n,w_0,s,\lambda} & = & \lambda\mbox{-}\sym{HWB}_{n,r} \circ \phi_{n,w_0,s} = \lambda \circ \sym{HWB}_{n,r} \circ \phi_{n,w_0,s}.
\end{eqnarray}
So for $\mathbf{x}\in\mathbb{F}_2^n$, 
\begin{eqnarray}\label{eqn-int-lam-hwb-map}
	\sym{IntHWB}_{n,w_0,s,\lambda}(\mathbf{x}) & = & \lambda(\sym{HWB}_{n,r}(\phi_{n,w_0,s}(\mathbf{x}))). 
\end{eqnarray}
One may note that the application of $\phi_{n,w_0,s}$ to $\mathbf{x}$ corresponds to a pre-processing of the input to $\lambda\mbox{-}\sym{HWB}_{n,r}$.

\paragraph{Efficiency.}
The parameters to the map $\sym{IntHWB}_{n,w_0,s,\lambda}$ are the integers $w_0\in \mathbb{Z}_{2^n}$, $s<n$ and the $r$-variable function $\lambda$. 
The number of bits required to store $w_0$ is $n$ and the number of bits required to store $s$ is $\lceil\log_2n\rceil$. 
Assuming that $\lambda$ is stored in its truth table representation, $\sym{IntHWB}_{n,w_0,s,\lambda}$ requires $n+\lceil\log_2n\rceil+2^r$ bits to be stored. 
Note that we consider $r=5$ and so $\sym{IntHWB}_{n,w_0,s,\lambda}$ has a very efficient space representation. 
Computing $\phi_{n,w_0,s}$ requires computing $\mathcal{B}_{n,w_0,s}$ twice and the reversal of an $n$-bit string. The computation of
$\mathcal{B}_{n,w_0,s}$ requires the computation of $\mathcal{I}_{n,w_0,s}$ (which requires an integer subtraction, optionally an integer addition
and a right shift operation to implement the quotient), four truncations to implement the modulo operations, one left shift operation to implement the multiplication
by $2^{n-s}$, two integer additions, a left shift and an increment to compute $2k+1$, one integer subtraction, and a single integer multiplication.
In a hardware implementation, $w_0$ and $s$ will be hardcoded into the circuit and $\lambda$ will be implemented as a small combinational circuit. The other
operations are simple arithmetic operations on $n$-bit quantities. In particular, for $n$ around 20, we do not expect the size of the circuit implementing 
$\sym{IntHWB}_{n,w_0,s,\lambda}$ to be too large; a rough estimate is about a few hundred gates.
So $\sym{IntHWB}_{n,w_0,s,\lambda}$ is a very efficiently implementable function. 

\begin{proposition}\label{prop-int-lambda-hwb-bal}
	Let $n\geq 2$, $w_0\in \mathbb{Z}_{2^n}$, $s,r<n$ be positive integers, and $\lambda$ be an $r$-variable function. Then $\sym{IntHWB}_{n,w_0,s,\lambda}$
	is balanced if and only if $\lambda$ is balanced. 
\end{proposition}
\begin{proof}
	Since $\phi_{n,w_0,s}$ is a bijection, $\phi_{n,w_0,s}\circ \lambda\mbox{-}\sym{HWB}_{n,r}$ is balanced if and only if 
	$\lambda\mbox{-}\sym{HWB}_{n,r}$ is balanced if and only if $\lambda$ is balanced. 
\end{proof}

The requirement is to choose $w_0$, $s$ and $\lambda$ in a manner so that $\sym{IntHWB}_{n,w_0,s,\lambda}$ has high nonlinearity. Since $\sym{IntHWB}_{n,w_0,s,\lambda}$ 
is constructed
using the composition operator, using~\eqref{eqn-WT-comp} the Walsh transform of $\sym{IntHWB}_{n,w_0,s,\lambda}$ can be expressed in terms of the Walsh transforms of
$\phi_{n,w_0,s}$, $\sym{HWB}_{n,r}$ and $\lambda$. The resulting expression, however, does not provide guidance on how to choose the parameters of
$\sym{IntHWB}_{n,w_0,s,\lambda}$ to ensure high nonlinearity. Further, we are also not aware of any other analytical method for ensuring that $\sym{IntHWB}_{n,w_0,s,\lambda}$
has high nonlinearity. In view of this, we decided to search for appropriate parameters so that $\sym{IntHWB}_{n,w_0,s,\lambda}$ has high nonlinearity. 
Letting $w_0\in \mathbb{Z}_{2^n}$, $s\leq \lfloor n/2\rfloor$ and $\lambda$ to be a balanced $r$-variable function make the size of the parameter space
$O(n2^n{2^r\choose 2^{r-1}})$. For each selection of parameters in this space, it is required to construct the function $\sym{IntHWB}_{n,w_0,s,\lambda}$ and compute
its nonlinearity. This requires $O(n2^n)$ time. So the total time for the search becomes $O(n^22^{2n}{2^r\choose 2^{r-1}})$. This is computationally infeasible. So we decided
to fix $r=5$ and consider the functions $\lambda_{5,i}$ corresponding to the values of $n$ given by Proposition~\ref{prop-lambda-hwb-req5}. This reduces the
search time to $O(n^22^{2n})$. For $n=13,\ldots,20$ we were able to carry out this search. The search algorithm is given in Algorithm~\ref{algo-srch1}.

\begin{algorithm}[h]
\SetAlgorithmName{Algorithm}{}
\DontPrintSemicolon
	\KwInput{$n$, $\mathcal{L}$, where $\mathcal{L}$ is the list of $\lambda_{5,i}$ corresponding to $n$ as given in Proposition~\ref{prop-lambda-hwb-req5}}
	\KwOutput{A list $\mathcal{P}$ of triplets $(\lambda,s,w_0)$.}

	$\sym{maxnl}\leftarrow 0$; $\mathcal{P}\leftarrow \emptyset$ \\
        \For{$\lambda\in \mathcal{L}$} {
		\For{$s$ in $\{1,\ldots,\lfloor n/2\rfloor$} { 
			\For{$w_0$ in $\mathbb{Z}_{2^n}$} { 
				let $f=\sym{IntHWB}_{n,w_0,s,\lambda}$ \\
				compute $\sym{nl}(f)$ and $\sym{deg}(f)$ \\
				\If{$\sym{deg}(f) = n-1$ and $\sym{maxnl} < \sym{nl}(f)$} {
					$\sym{maxnl} \leftarrow \sym{nl}(f)$; $\mathcal{P}\leftarrow \{(\lambda,s,w_0)\}$
				}
				\Else {
					\If{$\sym{deg}(f) = n-1$ and $\sym{maxnl}=\sym{nl}(f)$} {
						$\mathcal{P}\leftarrow \mathcal{P} \cup \{(\lambda,s,w_0)\}$
					}
				}
			}
		}
        }
        return $\mathcal{P}$
        \caption{The search procedure for $\sym{IntHWB}_{n,w_0,s,\lambda}$. \label{algo-srch1}}
\end{algorithm}
%\begin{proposition}\label{prop-algo-srch1-bal}
%	Suppose $\mathcal{P}$ is returned by Algorithm~\ref{algo-srch1} on input $n$. For each $(\lambda,s,w_0)\in \mathcal{P}$, the function
%	$\sym{IntHWB}_{n,w_0,s,\lambda}$ is a balanced $n$-variable function having degree equal to $n-1$.
%\end{proposition}
The results of running Algorithm~\ref{algo-srch1} for $n=13,\ldots,20$ are stated in the following proposition.
\begin{proposition}\label{prop-int-lambda-hwb}
	For $n=13,\ldots,20$ and $\lambda$ is one of $\lambda_{5,i}$ given by Proposition~\ref{prop-lambda-hwb-req5}, the maximum nonlinearities achieved
	by $\sym{IntHWB}_{n,w_0,s,\lambda}$ are as follows.
	%\begin{enumerate}
	\begin{compactenum}
		\item $n=13$: for $s=4$, $w_0=254$, $\sym{nl}(\sym{IntHWB}_{n,w_0,s,\lambda_{5,2}})=3952$.
		\item $n=14$: for $s=5$, $w_0=13090$, $\sym{nl}(\sym{IntHWB}_{n,w_0,s,\lambda_{5,4}})=7974$.
		\item $n=15$: for $s=7$, $w_0=21272$, $\sym{nl}(\sym{IntHWB}_{n,w_0,s,\lambda_{5,2}})=16062$.
		\item $n=16$: \\
			for $s=4$, $w_0=16699$, $\sym{nl}(\sym{IntHWB}_{n,w_0,s,\lambda_{5,5}})=32290$; \\
		        for $s=4$, $w_0=27429$, $\sym{nl}(\sym{IntHWB}_{n,w_0,s,\lambda_{5,5}})=32290$.
		\item $n=17$: for $s=4$, $w_0=105883$, $\sym{nl}(\sym{IntHWB}_{n,w_0,s,\lambda_{5,1}})=64834$.
		\item $n=18$: for $s=5$, $w_0=118924$, $\sym{nl}(\sym{IntHWB}_{n,w_0,s,\lambda_{5,7}})=130042$.
		\item $n=19$: for $s=5$, $w_0=200085$, $\sym{nl}(\sym{IntHWB}_{n,w_0,s,\lambda_{5,9}})=260606$.
		\item $n=20$: for $s=5$, $w_0=353518$, $\sym{nl}(\sym{IntHWB}_{n,w_0,s,\lambda_{5,12}})=522046$.
	\end{compactenum}
	%\end{enumerate}
\end{proposition}

%\subsection{Resistance to Algebraic Attacks \label{subsec-aa}}
%For computing algebraic immunities we used the Boolean function 
%library\footnote{\url{https://doc.sagemath.org/html/en/reference/cryptography/sage/crypto/boolean_function.html\#sage.crypto.boolean_function.BooleanFunction.annihilator}} of 
%the SageMath software.
For $n=13,\ldots,19$, the algebraic immunities of the functions given in Proposition~\ref{prop-int-lambda-hwb} could be computed on our servers, but
for $n=20$, the process exited abnormally and did not return the value of AI. The values of AI for $n=13,\ldots,19$ are stated in the following proposition.
\begin{proposition}\label{prop-int-lambda-hwb-AI}
	The algebraic immunities of the functions in Proposition~\ref{prop-int-lambda-hwb} are as follows.
        %\begin{enumerate}
	\begin{compactenum}
                \item $n=13$: for $s=4$, $w_0=254$, $\sym{AI}(\sym{IntHWB}_{n,w_0,s,\lambda_{5,2}})=6$.
                \item $n=14$: for $s=5$, $w_0=13090$, $\sym{AI}(\sym{IntHWB}_{n,w_0,s,\lambda_{5,4}})=7$.
                \item $n=15$: for $s=7$, $w_0=21272$, $\sym{AI}(\sym{IntHWB}_{n,w_0,s,\lambda_{5,2}})=7$.
                \item $n=16$: \\
                        for $s=4$, $w_0=16699$, $\sym{AI}(\sym{IntHWB}_{n,w_0,s,\lambda_{5,5}})=8$; \\
                        for $s=4$, $w_0=27429$, $\sym{AI}(\sym{IntHWB}_{n,w_0,s,\lambda_{5,5}})=8$.
                \item $n=17$: for $s=4$, $w_0=105883$, $\sym{AI}(\sym{IntHWB}_{n,w_0,s,\lambda_{5,1}})=9$.
                \item $n=18$: for $s=5$, $w_0=118924$, $\sym{AI}(\sym{IntHWB}_{n,w_0,s,\lambda_{5,7}})=9$.
		\item $n=19$: for $s=5$, $w_0=200085$, $\sym{nl}(\sym{IntHWB}_{n,w_0,s,\lambda_{5,9}})=9$.
	\end{compactenum}
        %\end{enumerate}
\end{proposition}
Note that except for $n=13,15$ and $19$, in all other cases the algebraic immunities are optimal, and for $n=13,15$ and $19$, the algebraic immunities are
one less than the optimal. We conjecture that the value of AI for the function in Proposition~\ref{prop-int-lambda-hwb} for $n=20$ is 10. 
This is based on our further study of algebraic immunities as discussed below.

To further understand the algebraic immunities of the functions in the class $\sym{IntHWB}_{n,w_0,s,\lambda}$, we conducted some more experiments. 
For $n=13,\ldots,19$, we fixed $\lambda$ and $s$ as in Proposition~\ref{prop-int-lambda-hwb-AI} and for 100 randomly chosen values of $w_0$, we constructed the 
function $\sym{IntHWB}_{n,w_0,s,\lambda}$ and computed its nonlinearity and algebraic immunity.
For $n=14,16$ and $18$, in all the 100 cases the algebraic immunities came out to be $n/2$, i.e. optimal. For $n=13,15,17$ and $19$, in all the 100 cases the algebraic 
immunities came out to be either $\lfloor n/2\rfloor$ or $\lceil n/2\rceil$. Letting $a_1$ and $a_2$ to be the number of cases where the algebraic immunities came out to be
$\lfloor n/2\rfloor$ and $\lceil n/2\rceil$ respectively, we obtained $(a_1,a_2)=(70,30), (65,35), (73,27), (62,38)$ for $n=13,15, 17$ and $19$ respectively.
So the experiments provide evidence that for even $n$ functions in the class $\sym{IntHWB}_{n,w_0,s,\lambda}$ have optimal algebraic immunity, while for odd $n$,
functions in the class $\sym{IntHWB}_{n,w_0,s,\lambda}$ have either optimal or almost optimal algebraic immunity, with optimal algebraic immunity occuring for about 30\%
or more of the cases. 

For $n=17$, the function in Proposition~\ref{prop-int-lambda-hwb} has optimal algebraic immunity. For $n=13,15$ and $19$, the functions
in Proposition~\ref{prop-int-lambda-hwb} have algebraic immunity one less than the optimal. From the results of our above mentioned experiments with 100 random values 
of $w_0$, we provide examples of functions for $n=13,15$ and $19$ with optimal algebraic immunity.
\begin{example} \label{ex-optAI-odd-n} \ 
\begin{itemize}
	\item $n=13$: for $s=3$, $w_0=3204$, \\ $\sym{nl}(\sym{IntHWB}_{n,w_0,s,\lambda_{5,7}})=3950$, $\sym{AI}(\sym{IntHWB}_{n,w_0,s,\lambda_{5,7}})=7$.
	\item $n=15$: for $s=4$, $w_0=51$, \\ $\sym{nl}(\sym{IntHWB}_{n,w_0,s,\lambda_{5,7}})=16036$, $\sym{AI}(\sym{IntHWB}_{n,w_0,s,\lambda_{5,7}})=8$.
	\item $n=19$: for $s=5$, $w_0=471438$, \\ $\sym{nl}(\sym{IntHWB}_{n,w_0,s,\lambda_{5,9}})=260502$, $\sym{AI}(\sym{IntHWB}_{n,w_0,s,\lambda_{5,7}})=10$.
\end{itemize}
\end{example}
Note that for $n=13$, the nonlinearity of the above example is 3950, while the maximum nonlinearity reported in Proposition~\ref{prop-int-lambda-hwb} is 3952. For
$n=15$, the nonlinearity of the above example is 16036, while the maximum nonlinearity reported in Proposition~\ref{prop-int-lambda-hwb} is 16062. 
For $n=19$, the nonlinearity of the above example is 260502, while the maximum nonlinearity reported in Proposition~\ref{prop-int-lambda-hwb} is 260606.
So for $n=13,15$ and $19$, optimal AI can be obtained with a small decrease in nonlinearity.

To assess the resistance of the class of functions to fast algebraic attacks, we computed the FAA-profile for the functions given in Proposition~\ref{prop-int-lambda-hwb} 
for $n=13,14,15$ and $16$ and also for the functions in Example~\ref{ex-optAI-odd-n}. These are given below.
\begin{itemize}
	\item FAA-profile for $\sym{IntHWB}_{n,w_0,s,\lambda_{5,2}}$ with $n=13$, $s=4$, $w_0=254$: \\ \quad $(1,11), (2,9), (3,9), (4,7), (5,7)$.
	\item FAA-profile for $\sym{IntHWB}_{n,w_0,s,\lambda_{5,7}}$ with $n=13$, $s=3$, $w_0=3204$: \\ \quad $(1,10), (2,9), (3,9), (4,7), (5,7), (6,6)$.
	\item FAA-profile for $\sym{IntHWB}_{n,w_0,s,\lambda_{5,4}}$ with $n=14$, $s=5$, $w_0=13090$: \\ \quad $(1,11), (2,11), (3,10), (4,8), (5,7), (6,7)$.
	\item FAA-profile for $\sym{IntHWB}_{n,w_0,s,\lambda_{5,2}})$ with $n=15$, $s=7$, $w_0=21272$: \\ \quad $(1,13), (2,11), (3,11), (4,9), (5,9), (6,7)$.
	\item FAA-profile for $\sym{IntHWB}_{n,w_0,s,\lambda_{5,7}}$ with $n=15$, $s=4$, $w_0=51$: \\ \quad $(1,13), (2,11), (3,10), (4,9), (5,8), (6,7), (7,7)$.
	\item FAA-profile for $\sym{IntHWB}_{n,w_0,s,\lambda_{5,5}}$ with $n=16$, $s=4$, $w_0=16699$: \\ \quad $(1,13), (2,12), (3,11), (4,10), (5,9), (6,8), (7,7)$.
	\item FAA-profile for $\sym{IntHWB}_{n,w_0,s,\lambda_{5,5}}$ with $n=16$, $s=4$, $w_0=27429$: \\ \quad $(1,13), (2,12), (3,11), (4,10), (5,9), (6,8), (7,7)$.
\end{itemize}
We find that almost perfect FAA-profile is achieved in all cases. 
Consequently, for all such functions $f$, $\sym{FAI}(f)\geq n-1$. This indicates good resistance of these functions to fast algebraic attacks.

For $n=17,\ldots,20$, due to high memory requirement, it was not possible to compute the complete FAA-profiles for the functions in Proposition~\ref{prop-int-lambda-hwb}
and Example~\ref{ex-optAI-odd-n}.
Below we provide the partial FAA-profiles that could be computed.
\begin{itemize}
	\item partial FAA-profile for $\sym{IntHWB}_{n,w_0,s,\lambda_{5,1}}$ with $n=17$, $s=4$, $w_0=105883$: \\ $(1,14), (2,14), (3,13), (4,12), (5,11)$.
	\item partial FAA-profile for $\sym{IntHWB}_{n,w_0,s,\lambda_{5,7}}$ with $n=18$, $s=5$, $w_0=118924$: \\ $(1,15), (2,15), (3,13), (4,12)$.
	\item partial FAA-profile for $\sym{IntHWB}_{n,w_0,s,\lambda_{5,9}}$ with $n=19$, $s=5$, $w_0=200085$: \\ $(1,16), (2,15), (3,14)$.
	\item partial FAA-profile for $\sym{IntHWB}_{n,w_0,s,\lambda_{5,9}}$ with $n=19$, $s=5$, $w_0=471438$: \\ $(1,17), (2,15), (3,15)$.
	\item partial FAA-profile for $\sym{IntHWB}_{n,w_0,s,\lambda_{5,12}}$ with $n=20$, $s=5$, $w_0=353518$: \\ $(1,17), (2,16), (3,15)$.
\end{itemize}
We observe that in all cases for $(e,d)$ in the above partial FAA-profiles, the relation $e+d\geq n-2$ holds and we conjecture that for any of these functions $f$,
the relation $\sym{FAI}(f)\geq n-1$ hold.

\begin{remark}\label{rem-aa-inthwb}
	From the experimental results we observe that for all the $n$-variable functions $f$ of the type $\sym{IntHWB}$, for which we were able to compute the algebraic immunities 
	and the FAA-profiles, we have $\sym{AI}(f)\geq \lfloor n/2\rfloor$, and $\sym{FAI}(f)\geq n-1$. Further, $\sym{AI}(f)=\lceil n/2\rceil$ in several of the cases. This 
	suggests that functions of the type $\sym{IntHWB}$ provide good resistance to algebraic and fast algebraic attacks. 
\end{remark}

\subsection{Comparison \label{subsec-comp-int-lambda-hwb}}
Among the previously known efficiently implementable functions HWB, GHWB \cite{DBLP:journals/tit/Carlet22} and the cyclic weightwise function~\cite{MO24,cryptoeprint:2024/2022}, 
the nonlinearities reported in Table~7 of~\cite{MO24} and Table~5 of~\cite{cryptoeprint:2024/2022} are the highest.
So we compare the nonlinearities reported in Table~7 of~\cite{MO24} and Table~5 of~\cite{cryptoeprint:2024/2022} with those of $\sym{IntHWB}_{n,w_0,s,\lambda}$. To provide 
context, we also 
compare to the nonlinearites of the CF functions (even though the CF functions are not efficiently implementable) as well as to the values of the covering radius bound. 
We constructed the CF functions using the primitive polynomials in Appendix~\ref{app-prim} and
then computed their nonlinearities. For $n=13$, the nonlinearity of the CF function that we obtained is higher than the nonlinearity reported in~\cite{DBLP:journals/tit/Carlet22}.
This is not surprising since the actual function and hence the value of the nonlinearity depends upon the actual primitive polynomial that is used.
\begin{table}
	\centering
	%\begin{center}
		{\small
	\begin{tabular}{|c||r|r||r|r||r|r||r|r||r|r||}
		\cline{2-11}
		\multicolumn{1}{c}{} 
		& \multicolumn{2}{|c|}{Table~7 of~\cite{MO24}} 
		& \multicolumn{2}{|c|}{Table~5 of~\cite{cryptoeprint:2024/2022}} 
		& \multicolumn{2}{|c|}{Proposition~\ref{prop-int-lambda-hwb}} 
		& \multicolumn{2}{|c|}{CF~\cite{DBLP:conf/asiacrypt/CarletF08}} 
		& \multicolumn{2}{|c|}{cov rad bnd} \\ \hline
			$n$ & 
		\multicolumn{1}{|c|}{$\sym{nl}$} & \multicolumn{1}{|c|}{$\sym{LLB}$} &
		\multicolumn{1}{|c|}{$\sym{nl}$} & \multicolumn{1}{|c|}{$\sym{LLB}$} &
		\multicolumn{1}{|c|}{$\sym{nl}$} & \multicolumn{1}{|c|}{$\sym{LLB}$} &
		\multicolumn{1}{|c|}{$\sym{nl}$} & \multicolumn{1}{|c|}{$\sym{LLB}$} &
		\multicolumn{1}{|c|}{$\sym{CRB}_n$} & \multicolumn{1}{|c|}{$\sym{LCRB}_n$} \\ \hline
			$13$ & $3862$  & $-5.13$ & --      & --     & $3952$ & $-5.83$ & $3988$ & $-6.25$ & $4051$ & $-7.51$ \\ \hline 
			$14$ & $7816$  & $-5.45$ & $7842$  & $-5.55$ & $7974$ & $-6.23$ & $8072$ & $-7.09$ & $8128$ & $-8.00$ \\ \hline
			$15$ & $15748$ & $-5.69$ & --      & --      & $16062$ & $-6.67$ & $16212$ & $-7.57$ & $16294$ & $-8.51$ \\ \hline
			$16$ & $31616$ & $-5.83$ & $31680$ & $-5.91$ & $32290$ & $-7.10$ & $32530$ & $-8.11$ & $32640$ & $-9.00$ \\ \hline
			$17$ & --      & --      & --      & --      & $64834$ & $-7.54$ & $65210$ & $-8.65$ & $65355$ & $-9.50$ \\ \hline
			$18$ & --      & --      & --      & --      & $130042$ & $-7.99$ & $130594$ & $-9.10$ & $130816$ & $-10.00$ \\ \hline
			$19$ & --      & --      & --      & --      & $260606$ & $-8.41$ & $261294$ & $-9.27$ & $261782$ & $-10.50$ \\ \hline
			$20$ & --      & --      & --      & --      & $522046$ & $-8.87$ & $523234$ & $-9.96$ & $523776$ & $-11.00$ \\ \hline
	\end{tabular}
		\caption{Comparison of nonlinearities achieved by $\sym{IntHWB}_{n,w_0,s,\lambda}$ with Table~7 of~\cite{MO24}, Table~5 of~\cite{cryptoeprint:2024/2022},
		the CF functions and the covering radius bound. \label{tab-nl-comp-int}}
	%\end{center}
%\end{table}
%\begin{table}

	\centering
%\begin{center}
        \begin{tabular}{|c|c|c|c|c|}
                \hline
		$n$ & Table~7 of~\cite{MO24} 
		    & Tables~6 and~7 of~\cite{cryptoeprint:2024/2022} 
		    & Proposition~\ref{prop-int-lambda-hwb-AI} 
		    & CF~\cite{DBLP:conf/asiacrypt/CarletF08} \\ \hline
		13  & (12,6) & --     & (12,6) & (12,7) \\ \hline
		14  & (12,6) & (13,6) & (13,7) & (13,7) \\ \hline
		15  & (14,6) & --     & (14,7) & (14,8) \\ \hline
		16  & (14,7) & (15,7) & (15,8) & (15,8) \\ \hline
		17  & --     & --     & (16,9) & (16,9) \\ \hline
		18  & --     & --     & (17,9) & (17,9) \\ \hline
		19  & --     & --     & (18,9) & (18,10) \\ \hline
        \end{tabular}
	\caption{Comparison of degrees and algebraic immunities of $\sym{IntHWB}_{n,w_0,s,\lambda}$ functions with Table~7 of~\cite{MO24}, 
	Tables~6 and~7 of~\cite{cryptoeprint:2024/2022} and the CF functions. \label{tab-AI-comp-int} }
	}
%\end{center}
\end{table}
The comparison of nonlinearities is shown in Table~\ref{tab-nl-comp-int}. 
The comparison of degrees and algebraic immunities are shown in Table~\ref{tab-AI-comp-int}. 
Each entry of Table~\ref{tab-AI-comp-int} is of the form $(d,a)$, where $d$ is the degree and $a$ is the algebraic immunity.
We note that the nonlinearities of the $\sym{IntHWB}$ functions reported in Proposition~\ref{prop-int-lambda-hwb} are higher than the nonlinearities reported in 
Table~7 of~\cite{MO24} and Table~5 of~\cite{cryptoeprint:2024/2022}. For $n=13$, the algebraic immunity of the $\sym{IntHWB}$ function given by 
Proposition~\ref{prop-int-lambda-hwb-AI} is equal to the algebraic immunity of the function reported in Table~7 of~\cite{MO24}. 

The $\sym{LLB}$'s of $\sym{IntHWB}$ functions are about 1.5 bits more than the $\sym{LLB}$'s of CF functions. While this may seem like a disadvantage, it
is not actually so. Suppose a target value of $\sym{LLB}$ is fixed and the value is achieved by CF functions for a particular value
of $n$. By choosing a higher value of $n$, the same value of $\sym{LLB}$ can be also be achieved by $\sym{IntHWB}$ functions. For example, choosing
$n=19$ we obtain a value of $\sym{LLB}$ which is lower than the value of the CF function for $n=16$. Since $\sym{IntHWB}$ for $n=19$ is very efficiently 
implementable, whereas the CF function with $n=16$ is not, there is no loss in increasing the number of variables.

\subsection{Examples of $\sym{IntHWB}_{n,w_0,s,\lambda}$ for $n=21$ to $30$ \label{sec-neq21to30}}
It becomes very time consuming to run Algorithm~\ref{algo-srch1} for $n$ greater than $20$. To obtain an idea of the nonlinearity achieved by 
$\sym{IntHWB}_{n,w_0,s,\lambda}$ for higher values of $n$ we conducted some experiments. We fixed $s=5$ and $\lambda=\lambda_{5,7}$ and constructed
$\sym{IntHWB}_{n,w_0,s,\lambda}$ for a number of random choices of $w_0$. For $n=21,\ldots,24$, we chose 10000 values for $w_0$, while for $n=25,\ldots,30$, we
chose 1000 values for $w_0$. For each $n=21,\ldots,30$, in the following example, we report the maximum nonlinearity that was achieved. 
\begin{example}\label{ex-hi-n}\  
\begin{itemize}
        \item $n=21$, $s=5$, $w_0=1948971$: $\sym{nl}(\sym{IntHWB}_{n,w_0,s,\lambda_{5,7}})=1045280$. 
        \item $n=22$, $s=5$, $w_0=223972$: $\sym{nl}(\sym{IntHWB}_{n,w_0,s,\lambda_{5,7}})=2092280$. 
        \item $n=23$, $s=5$, $w_0=2179192$: $\sym{nl}(\sym{IntHWB}_{n,w_0,s,\lambda_{5,7}})=4187200$. 
        \item $n=24$, $s=5$, $w_0=11878200$: $\sym{nl}(\sym{IntHWB}_{n,w_0,s,\lambda_{5,7}})=8378102$. 
        \item $n=25$, $s=5$, $w_0=17211712$: $\sym{nl}(\sym{IntHWB}_{n,w_0,s,\lambda_{5,7}})=16761306$. 
        \item $n=26$, $s=5$, $w_0=45478445$: $\sym{nl}(\sym{IntHWB}_{n,w_0,s,\lambda_{5,7}})=33530292$. 
        \item $n=27$, $s=5$, $w_0=67070690$: $\sym{nl}(\sym{IntHWB}_{n,w_0,s,\lambda_{5,7}})=67070690$. 
        \item $n=28$, $s=5$, $w_0=95163654$: $\sym{nl}(\sym{IntHWB}_{n,w_0,s,\lambda_{5,7}})=134157910$. 
        \item $n=29$, $s=5$, $w_0=224553125$: $\sym{nl}(\sym{IntHWB}_{n,w_0,s,\lambda_{5,7}})=268332760$. 
        \item $n=30$, $s=5$, $w_0=378168951$: $\sym{nl}(\sym{IntHWB}_{n,w_0,s,\lambda_{5,7}})=536691884$. 
\end{itemize}
\end{example}
In Table~\ref{tab-hi-n}, we compare the nonlinearities in Example~\ref{ex-hi-n} with those of the CF-function. Note that for $n=21,\ldots,30$, 
even though we were able to explore a very limited portion of the parameter space of $\sym{IntHWB}$ functions, the nonlinearities and the values of $\sym{LLB}$
that are achieved compare quite well to the corresponding values of the CF functions. In particular, the values of $\sym{LLB}$ for the $\sym{IntHWB}$ functions
is at most about $2$ more than those of the CF functions. As explained in Section~\ref{subsec-comp-int-lambda-hwb}, the main advantage of $\sym{IntHWB}$ functions is their
very efficient implementation. So a target value of $\sym{LLB}$ can be cheaply achieved by increasing the value of $n$. While a CF function would achieve the same
value of $\sym{LLB}$ for a smaller value of $n$, it would be much more efficient to implement an $\sym{IntHWB}$ function with a higher value of $n$.
Further, based on our experiments for $n=13$ to $n=20$ we conjecture that even for $n>20$, the $\sym{IntHWB}$ functions provide
good resistance to algebraic attacks (see Remark~\ref{rem-aa-inthwb}).
\begin{table}
	\centering
	{\small
	\begin{tabular}{|c||r|r||r|r||r|r||}
		\cline{2-7}
		\multicolumn{1}{c}{} & \multicolumn{2}{|c|}{Example~\ref{ex-hi-n}} 
				& \multicolumn{2}{|c|}{CF~\cite{DBLP:conf/asiacrypt/CarletF08}} 
				& \multicolumn{2}{|c|}{cov rad bnd} \\ \hline
			$n$ & 
			\multicolumn{1}{|c|}{$\sym{nl}$} & \multicolumn{1}{|c|}{$\sym{LLB}$} &
			\multicolumn{1}{|c|}{$\sym{nl}$} & \multicolumn{1}{|c|}{$\sym{LLB}$} &
			\multicolumn{1}{|c|}{$\sym{CRB}_n$} & \multicolumn{1}{|c|}{$\sym{LCRB}_n$} \\ \hline
			$21$ & $1045280$ &  $-9.31$ &     $1046846$ & $-10.24$ &    $1047852$ & $-11.50$ \\ \hline
			$22$ & $2092280$ &  $-9.75$    &  $2094936$ & $-10.89$ &    $2096128$ & $-12.00$ \\ \hline
			$23$ & $4187200$ &  $-10.21$   &  $4190834$ & $-11.24$ &    $4192856$ & $-12.50$ \\ \hline
			$24$ & $8378102$ &  $-10.64$   &  $8383446$ & $-11.67$ &    $8386560$ & $-13.00$ \\ \hline
			$25$ & $16761306$ & $-11.04$   &  $16769938$ & $-12.17$ &   $16774320$ & $-13.50$ \\ \hline
			$26$ & $33530292$ & $-11.44$   &  $33545384$ & $-12.86$ &   $33550336$ & $-14.00$ \\ \hline
			$27$ & $67070690$ & $-11.78$   &  $67097318$ & $-13.50$ &   $67103072$ & $-14.50$ \\ \hline
			$28$ & $134157910$ & $-12.13$  &  $134201202$ & $-13.99$ &  $134209536$ & $-15.00$ \\ \hline
			$29$ & $268332760$ & $-12.35$  &  $268409892$ & $-14.36$ &  $268423871$ & $-15.50$ \\ \hline
			$30$ & $536691884$ & $-12.55$  &  $536833704$ & $-14.82$ &  $536854528$ & $-16.00$ \\ \hline
	\end{tabular}
		\caption{Comparison of nonlinearities achieved by the functions in Example~\ref{ex-hi-n} with those of CF functions and the covering radius bound. \label{tab-hi-n}}
		}
\end{table}
%\section{Construction of $\lambda$-HWB Functions \label{sec-lambdaHWB}}
%\section{Construction of Interval $\lambda$-HWB Functions \label{sec-int-lambdaHWB}}

\section{Conclusion \label{sec-conclu} }
We provided constructions of Boolean functions which are good solutions to the ``the big single-output Boolean problem'' proposed in~\cite{DBLP:journals/tit/Carlet22}. 
The functions are built using simple arithmetic operations leading to these functions being efficient to implement. 

For the functions that we propose we provide experimental results on the nonlinearity and algebraic resistance. A theoretical direction of work would be 
to prove results on nonlinearity and algebraic resistance for these functions. One major problem with doing this is that there are no good
mathematical techniques for analysing nonlinearity and algebraic resistance of functions built using a combination of integer arithmetic and arithmetic 
over $\mathbb{F}_2$. We hope that there will be future research focus on developing such techniques.

Another possible direction of work would be to extend our approach to vectorial functions. The ability to simultaneously produce several bits (instead of one) will lead to
higher speed of keystream generation. On the other hand, producing more bits also provides the adversary with more information. For a vectorial function, it will
be required to consider the nonlinearity and algebraic resistance of all non-zero component functions (i.e., linear combinations of the coordinate functions). The
mathematical challenge is to ensure that each non-zero component function provides sufficient resistance to attacks. This makes the problem more difficult than
the construction of Boolean functions. Again, we hope this problem will be addressed in the future.

\section*{Acknowledgement} 
We thank Pierrick M\'{e}aux for his comments on an earlier version of the paper. Deng Tang provided us with a program written by 
Simon Fischer which we have used for computing fast algebraic immunity. We thank both of them.

%\bibliographystyle{plain}
%\bibliography{gen-hwb}

%\newpage 
\appendix

\section{Primitive Polynomials Used to Construct CF Functions \label{app-prim}}

For $n=13$ to $30$, the following primitive polynomials were used in the construction of the CF functions.
\begin{tabbing}
\ \ \ \ \= \kill
\> $x^{13} \oplus x^4 \oplus x^3 \oplus x \oplus 1$ \\
\> $x^{14} \oplus x^{12} x^{11} \oplus x \oplus 1$ \\
\> $x^{15} \oplus x \oplus 1$ \\
\> $x^{16} \oplus x^5 \oplus x^3 \oplus x^2 \oplus 1$ \\
\> $x^{17} \oplus x^3 \oplus 1$ \\
\> $x^{18} \oplus x^7 \oplus 1$ \\
\> $x^{19} \oplus x^6 \oplus x^5 \oplus x \oplus 1$ \\
\> $x^{20} \oplus x^3 \oplus 1$ \\
\> $x^{21} \oplus x^2 \oplus 1$ \\
\> $x^{22} \oplus x \oplus 1$ \\
\> $x^{23} \oplus x^5 \oplus 1$ \\
\> $x^{24} \oplus x^4 \oplus x^3 \oplus x \oplus 1$ \\
\> $x^{25} \oplus x^3 \oplus 1$ \\
\> $x^{26} \oplus x^6 \oplus x^2 \oplus x^1 \oplus 1$ \\
\> $x^{27} \oplus x^5 \oplus x^2 \oplus x^1 \oplus 1$ \\
\> $x^{28} \oplus x^3 \oplus 1$ \\
\> $x^{29} \oplus x^2 \oplus 1$ \\
\> $x^{30} \oplus x^{23} \oplus x^2 \oplus x^1 \oplus 1$
\end{tabbing}

\section{ANFs of $\lambda_{5,i}$, $i=1,\ldots,12$ \label{app-ANF}}

{\tiny
\begin{eqnarray*}
	\lefteqn{\lambda_{5,1}(X_1,X_2,X_3,X_4,X_5)} \\
	& = & X_1 X_2 X_3 \oplus X_1 X_2 X_4 \oplus X_1 X_2 \oplus X_1 X_3 X_4 X_5 \oplus X_1 X_3 \oplus X_1 X_5 \oplus X_1 \oplus X_3 X_5 \\
	& & \oplus X_4 X_5 \oplus X_4 \oplus X_5 \oplus 1 \\ % 1884850941
	\lefteqn{\lambda_{5,2}(X_1,X_2,X_3,X_4,X_5)} 
	\\ & = & X_1 X_2 X_3 X_5 \oplus X_1 X_2 X_3 \oplus X_1 X_2 X_5 \oplus X_1 X_3 X_5 \oplus X_1 \oplus X_2 X_3 X_5 \oplus X_2 X_3 \oplus X_2 X_4 X_5 \\
	& & \oplus X_2 X_4 \oplus X_3 X_4 X_5 \oplus X_3 X_4 \oplus X_3 X_5 \oplus X_4 X_5 \oplus X_4 \oplus X_5 \oplus 1 \\ % 1920128533
%X_1 X_2 X_3 X_5 \oplus X_1 X_2 X_3 \oplus X_1 X_2 X_5 \oplus X_1 X_3 X_5 \oplus X_1 \oplus X_2 X_3 X_5 \oplus X_2 X_3 \oplus X_2 X_4 X_5 \oplus X_2 X_4 \oplus X_3 X_4 X_5 \oplus X_3 X_4 \oplus X_3 X_5 \oplus X_4 X_5 \oplus X_4 \oplus X_5 \\  % 2374838762
%X_1 X_2 X_3 \oplus X_1 X_2 X_4 \oplus X_1 X_2 \oplus X_1 X_3 X_4 X_5 \oplus X_1 X_3 \oplus X_1 X_5 \oplus X_1 \oplus X_3 X_5 \oplus X_4 X_5 \oplus X_4 \oplus X_5 \\ % 2410116354
	\lefteqn{\lambda_{5,3}(X_1,X_2,X_3,X_4,X_5)} \\
	& = &  X_1 X_2 X_3 X_5 \oplus X_1 X_2 X_5 \oplus X_1 X_3 X_5 \oplus X_1 X_3 \oplus X_1 X_4 X_5 \oplus X_1 \oplus X_2 X_3 X_4 X_5 \oplus X_2 X_4 X_5 \\
	& & \oplus X_2 X_5 \oplus X_2 \oplus X_3 X_4 X_5 \oplus X_3 X_4 \oplus X_3 \oplus X_4 X_5 \oplus X_4 \oplus 1 \\ % 195937993
	\lefteqn{\lambda_{5,4}(X_1,X_2,X_3,X_4,X_5)} \\
	& = &  X_1 X_2 X_3 X_4 \oplus X_1 X_2 X_5 \oplus X_1 X_2 \oplus X_1 X_3 X_4 X_5 \oplus X_1 X_4 \oplus X_1 \oplus X_2 X_3 X_4 \oplus X_2 X_3 \\
	& & \oplus X_2 X_5 \oplus X_2 \oplus X_3 X_4 X_5 \oplus X_3 X_5 \oplus X_3 \oplus 1 \\ % 798346017
%X_1 X_2 X_3 X_4 \oplus X_1 X_2 X_5 \oplus X_1 X_2 \oplus X_1 X_3 X_4 X_5 \oplus X_1 X_4 \oplus X_1 \oplus X_2 X_3 X_4 \oplus X_2 X_3 \oplus X_2 X_5 \oplus X_2 \oplus X_3 X_4 X_5 \oplus X_3 X_5 \oplus X_3 \\ % 3496621278
%X_1 X_2 X_3 X_5 \oplus X_1 X_2 X_5 \oplus X_1 X_3 X_5 \oplus X_1 X_3 \oplus X_1 X_4 X_5 \oplus X_1 \oplus X_2 X_3 X_4 X_5 \oplus X_2 X_4 X_5 \oplus X_2 X_5 \oplus X_2 \oplus X_3 X_4 X_5 \oplus X_3 X_4 \oplus X_3 \oplus X_4 X_5 \oplus X_4 \\ % 4099029302
%$ X_1 X_2 X_3 \oplus X_1 X_2 X_4 \oplus X_1 X_2 \oplus X_1 X_3 X_4 X_5 \oplus X_1 X_3 \oplus X_1 X_5 \oplus X_1 \oplus X_3 X_5 \oplus X_4 X_5 \oplus X_4 \oplus X_5 \oplus 1 $ % 1884850941
%$ X_1 X_2 X_3 X_5 \oplus X_1 X_2 X_3 \oplus X_1 X_2 X_5 \oplus X_1 X_3 X_5 \oplus X_1 \oplus X_2 X_3 X_5 \oplus X_2 X_3 \oplus X_2 X_4 X_5 \oplus X_2 X_4 \oplus X_3 X_4 X_5 \oplus X_3 X_4 \oplus X_3 X_5 \oplus X_4 X_5 \oplus X_4 \oplus X_5 \oplus 1 $ % 1920128533
%	\lambda_{5,5}(X_1,X_2,X_3,X_4,X_5) & = & X_1 X_2 X_3 X_5 \oplus X_1 X_2 X_3 \oplus X_1 X_2 X_5 \oplus X_1 X_3 X_5 \oplus X_1 \oplus X_2 X_3 X_5 \oplus X_2 X_3 \oplus X_2 X_4 X_5 \\
%	& & \oplus X_2 X_4 \oplus X_3 X_4 X_5 \oplus X_3 X_4 \oplus X_3 X_5 \oplus X_4 X_5 \oplus X_4 \oplus X_5 \\ % 2374838762
%	\lambda_{5,6}(X_1,X_2,X_3,X_4,X_5) & = &  X_1 X_2 X_3 \oplus X_1 X_2 X_4 \oplus X_1 X_2 \oplus X_1 X_3 X_4 X_5 \oplus X_1 X_3 \oplus X_1 X_5 \oplus X_1 \oplus X_3 X_5 \\
%	& & \oplus X_4 X_5 \oplus X_4 \oplus X_5 \\ % 2410116354
	\lefteqn{\lambda_{5,5}(X_1,X_2,X_3,X_4,X_5)} \\
	& = & X_1 X_2 X_3 X_5 \oplus X_1 X_2 X_3 \oplus X_1 X_2 X_5 \oplus X_1 X_3 X_5 \oplus X_1 X_5 \oplus X_1 \oplus X_2 X_3 X_4 \oplus X_2 X_4 X_5 \\
	& & \oplus X_2 X_4 \oplus X_3 X_4 X_5 \oplus X_3 X_4 \oplus X_3 X_5 \oplus X_4 X_5 \oplus X_4 \oplus X_5 \oplus 1 \\ % 416863957
	\lefteqn{\lambda_{5,6}(X_1,X_2,X_3,X_4,X_5)} \\
	& = & X_1 X_2 X_3 \oplus X_1 X_2 X_4 \oplus X_1 X_2 \oplus X_1 X_3 X_4 X_5 \oplus X_1 X_3 X_4 \oplus X_1 X_4 \oplus X_1 \oplus X_2 X_3 X_4 \\
	& & \oplus X_2 X_3 \oplus X_2 X_4 \oplus X_2 \oplus X_3 X_4 \oplus X_3 X_5 \oplus X_3 \oplus X_4 X_5 \oplus 1 \\ % 1884413857
%X_1 X_2 X_3 \oplus X_1 X_2 X_4 \oplus X_1 X_2 \oplus X_1 X_3 X_4 X_5 \oplus X_1 X_3 X_4 \oplus X_1 X_4 \oplus X_1 \oplus X_2 X_3 X_4 \oplus X_2 X_3 \oplus X_2 X_4 \oplus X_2 \oplus X_3 X_4 \oplus X_3 X_5 \oplus X_3 \oplus X_4 X_5 \\ % 2410553438
%X_1 X_2 X_3 X_5 \oplus X_1 X_2 X_3 \oplus X_1 X_2 X_5 \oplus X_1 X_3 X_5 \oplus X_1 X_5 \oplus X_1 \oplus X_2 X_3 X_4 \oplus X_2 X_4 X_5 \oplus X_2 X_4 \oplus X_3 X_4 X_5 \oplus X_3 X_4 \oplus X_3 X_5 \oplus X_4 X_5 \oplus X_4 \oplus X_5 \\ % 3878103338
%$ X_1 X_2 X_3 \oplus X_1 X_2 X_4 \oplus X_1 X_2 \oplus X_1 X_3 X_4 X_5 \oplus X_1 X_3 \oplus X_1 X_5 \oplus X_1 \oplus X_3 X_5 \oplus X_4 X_5 \oplus X_4 \oplus X_5 \oplus 1 $ % 1884850941
%$ X_1 X_2 X_3 X_5 \oplus X_1 X_2 X_3 \oplus X_1 X_2 X_5 \oplus X_1 X_3 X_5 \oplus X_1 \oplus X_2 X_3 X_5 \oplus X_2 X_3 \oplus X_2 X_4 X_5 \oplus X_2 X_4 \oplus X_3 X_4 X_5 \oplus X_3 X_4 \oplus X_3 X_5 \oplus X_4 X_5 \oplus X_4 \oplus X_5 \oplus 1 $ % 1920128533
%	\lambda_{5,9}(X_1,X_2,X_3,X_4,X_5) & = &  X_1 X_2 X_3 X_5 \oplus X_1 X_2 X_3 \oplus X_1 X_2 X_5 \oplus X_1 X_3 X_5 \oplus X_1 \oplus X_2 X_3 X_5 \oplus X_2 X_3 \oplus X_2 X_4 X_5 \\
%	& & \oplus X_2 X_4 \oplus X_3 X_4 X_5 \oplus X_3 X_4 \oplus X_3 X_5 \oplus X_4 X_5 \oplus X_4 \oplus X_5 \\ % 2374838762
%	\lambda_{5,10}(X_1,X_2,X_3,X_4,X_5) & = &  X_1 X_2 X_3 \oplus X_1 X_2 X_4 \oplus X_1 X_2 \oplus X_1 X_3 X_4 X_5 \oplus X_1 X_3 \oplus X_1 X_5 \oplus X_1 \oplus X_3 X_5 \\
%	& & \oplus X_4 X_5 \oplus X_4 \oplus X_5 \\ % 2410116354
	\lefteqn{\lambda_{5,7}(X_1,X_2,X_3,X_4,X_5)} \\
	& = & X_1 X_2 X_4 \oplus X_1 X_2 X_5 \oplus X_1 X_2 \oplus X_1 X_3 X_4 X_5 \oplus X_1 X_3 X_4 \oplus X_1 X_3 X_5 \oplus X_1 X_4 \oplus X_3 \\
	& & \oplus X_4 X_5 \oplus X_4 \oplus X_5 \oplus 1 \\ % 1917909639
	\lefteqn{\lambda_{5,8}(X_1,X_2,X_3,X_4,X_5)} \\
	& = & X_1 X_2 X_3 X_5 \oplus X_1 X_2 X_3 \oplus X_1 X_2 X_5 \oplus X_1 X_4 X_5 \oplus X_1 X_4 \oplus X_1 X_5 \oplus X_1 \oplus X_2 X_4 X_5 \\
	& & \oplus X_2 X_4 \oplus X_3 X_5 \oplus X_4 X_5 \oplus X_4 \oplus X_5 \oplus 1 \\ % 2021149909
%X_1 X_2 X_3 X_5 \oplus X_1 X_2 X_3 \oplus X_1 X_2 X_5 \oplus X_1 X_4 X_5 \oplus X_1 X_4 \oplus X_1 X_5 \oplus X_1 \oplus X_2 X_4 X_5 \oplus X_2 X_4  \oplus X_3 X_5 \oplus X_4 X_5 \oplus X_4 \oplus X_5 \\ % 2273817386
%\lambda_{5,8}(X_1,X_2,X_3,X_4,X_5) & = &  X_1 X_2 X_4 \oplus X_1 X_2 X_5 \oplus X_1 X_2 \oplus X_1 X_3 X_4 X_5 \oplus X_1 X_3 X_4 \oplus X_1 X_3 X_5 \oplus X_1 X_4 \oplus X_3 \oplus X_4 X_5 \oplus X_4 \oplus X_5 \\ % 2377057656
	\lefteqn{\lambda_{5,9}(X_1,X_2,X_3,X_4,X_5)} \\
	& = & X_1 X_2 X_3 X_4 \oplus X_1 X_2 X_3 \oplus X_1 X_2 X_4 X_5 \oplus X_1 X_2 X_4 \oplus X_1 X_2 \oplus X_1 X_3 X_4 \oplus X_1 X_5 \oplus X_1 \\
	& & \oplus X_2 X_3 X_4 X_5 \oplus X_2 X_4 \oplus X_2 \oplus X_3 X_4 \oplus X_3 X_5 \oplus X_4 \oplus X_5 \oplus 1 \\ % 389347985
	\lefteqn{\lambda_{5,10}(X_1,X_2,X_3,X_4,X_5)} \\
	& = &  X_1 X_2 X_3 X_4 \oplus X_1 X_2 X_3 \oplus X_1 X_2 X_4 X_5 \oplus X_1 X_2 X_5 \oplus X_1 X_3 X_4 \oplus X_1 X_4 X_5 \oplus X_1 \oplus X_2 X_3 X_4 X_5 \\
	& & \oplus X_2 X_4 X_5 \oplus X_2 X_4 \oplus X_3 X_4 \oplus X_3 X_5 \oplus X_4 X_5 \oplus 1 \\ % 1622002133
%X_1 X_2 X_3 X_4 \oplus X_1 X_2 X_3 \oplus X_1 X_2 X_4 X_5 \oplus X_1 X_2 X_5 \oplus X_1 X_3 X_4 \oplus X_1 X_4 X_5 \oplus X_1 \oplus X_2 X_3 X_4 X_5 \oplus X_2 X_4 X_5 \oplus X_2 X_4 \oplus X_3 X_4 \oplus X_3 X_5 \oplus X_4 X_5 \\ % 2672965162
%X_1 X_2 X_3 X_4 \oplus X_1 X_2 X_3 \oplus X_1 X_2 X_4 X_5 \oplus X_1 X_2 X_4 \oplus X_1 X_2 \oplus X_1 X_3 X_4 \oplus X_1 X_5 \oplus X_1 \oplus X_2 X_3 X_4 X_5 \oplus X_2 X_4 \oplus X_2 \oplus X_3 X_4 \oplus X_3 X_5 \oplus X_4 \oplus X_5 \\ % 3905619310
	\lefteqn{\lambda_{5,11}(X_1,X_2,X_3,X_4,X_5)} \\
	& = &  X_1 X_2 X_3 X_4 \oplus X_1 X_2 X_3 X_5 \oplus X_1 X_2 X_3 \oplus X_1 X_2 X_4 X_5 \oplus X_1 X_2 X_5 \oplus X_1 X_3 X_4 X_5 \oplus X_1 X_3 X_5 \oplus X_1 X_3 \\
	& & \oplus X_1 X_5 \oplus X_1 \oplus X_2 X_3 X_4 X_5 \oplus X_2 X_4 \oplus X_2 X_5 \oplus X_2 \oplus X_3 X_5 \oplus X_4 X_5 \oplus X_4 \oplus X_5 \\ % 596112710
	\lefteqn{\lambda_{5,12}(X_1,X_2,X_3,X_4,X_5)} \\
	& = &  X_1 X_2 X_3 X_4 \oplus X_1 X_2 X_3 X_5 \oplus X_1 X_2 X_4 X_5 \oplus X_1 X_3 X_4 X_5 \oplus X_1 X_3 X_5 \oplus X_1 X_3 \oplus X_1 X_4 X_5 \oplus X_1 X_4 \\
	& & \oplus X_1 \oplus X_2 X_3 X_4 X_5 \oplus X_2 X_3 \oplus X_2 X_5 \oplus X_2 \oplus X_3 X_4 X_5 \oplus X_4 X_5 \oplus X_4 \\ % 1927992070
%$ X_1 X_2 X_3 X_4 \oplus X_1 X_2 X_3 X_5 \oplus X_1 X_2 X_4 X_5 \oplus X_1 X_3 X_4 X_5 \oplus X_1 X_3 X_5 \oplus X_1 X_3 \oplus X_1 X_4 X_5 \oplus X_1 X_4 \oplus X_1 \oplus X_2 X_3 X_4 X_5 \oplus X_2 X_3 \oplus X_2 X_5 \oplus X_2 \oplus X_3 X_4 X_5 \oplus X_4 X_5 \oplus X_4 \oplus 1 $ % 2366975225
%$ X_1 X_2 X_3 X_4 \oplus X_1 X_2 X_3 X_5 \oplus X_1 X_2 X_3 \oplus X_1 X_2 X_4 X_5 \oplus X_1 X_2 X_5 \oplus X_1 X_3 X_4 X_5 \oplus X_1 X_3 X_5 \oplus X_1 X_3 \oplus X_1 X_5 \oplus X_1 \oplus X_2 X_3 X_4 X_5 \oplus X_2 X_4 \oplus X_2 X_5 \oplus X_2 \oplus X_3 X_5 \oplus X_4 X_5 \oplus X_4 \oplus X_5 \oplus 1 $ % 3698854585
\end{eqnarray*}
}

\end{document}